\documentstyle[12pt]{article}

\newcommand{\eqn}[1]{(\ref{#1})}

\newcommand{\complex}{{\bb C}} 
\newcommand{\complexs}{{\bbs C}} 
\newcommand{\zed}{{\bb Z}} 
\newcommand{\real}{{\bb R}} 
\newcommand{\zeds}{{\bbs Z}} 
\newcommand{\id}{{\bb I}} 
\newcommand{\alg}{{\cal A}} 

\def\norm#1{{\Vert#1\Vert}}
\newcommand{\NO}{\,\mbox{$\circ\atop\circ$}\,} 

\font\mybb=msbm10 at 12pt
\def\bb#1{\hbox{\mybb#1}}
\font\mybbs=msbm10 at 9pt
\def\bbs#1{\hbox{\mybbs#1}}

\def\nn{\nonumber}
\def\e{{\rm e}}
\def\semiprod{{\supset\!\!\!\!\!\!\!\times~}}
\def\slash{\!\!\!/}
\def\slashs{\!\!\!\!/}
\def\Dirac{{D\!\!\!\!/\,}}
\def\beq{\begin{equation}}
\def\eeq{\end{equation}}
\def\bea{\begin{eqnarray}}
\def\eea{\end{eqnarray}}
\def\bd{\begin{displaymath}}
\def\ed{\end{displaymath}}

\renewcommand{\theequation}{\thesection.\arabic{equation}}

\makeatletter
\newdimen\normalarrayskip              
\newdimen\minarrayskip                 
\normalarrayskip\baselineskip
\minarrayskip\jot
\newif\ifold             \oldtrue            \def\new{\oldfalse}
\def\arraymode{\ifold\relax\else\displaystyle\fi} 
\def\@arrayskip{\ifold\baselineskip\z@\lineskip\z@
     \else
     \baselineskip\minarrayskip\lineskip2\minarrayskip\fi}
\def\@arrayclassz{\ifcase \@lastchclass \@acolampacol \or
\@ampacol \or \or \or \@addamp \or
   \@acolampacol \or \@firstampfalse \@acol \fi
\edef\@preamble{\@preamble
  \ifcase \@chnum
     \hfil$\relax\arraymode\@sharp$\hfil
     \or $\relax\arraymode\@sharp$\hfil
     \or \hfil$\relax\arraymode\@sharp$\fi}}
\def\@array[#1]#2{\setbox\@arstrutbox=\hbox{\vrule
     height\arraystretch \ht\strutbox
     depth\arraystretch \dp\strutbox
     width\z@}\@mkpream{#2}\edef\@preamble{\halign \noexpand\@halignto
\bgroup \tabskip\z@ \@arstrut \@preamble \tabskip\z@ \cr}%
\let\@startpbox\@@startpbox \let\@endpbox\@@endpbox
  \if #1t\vtop \else \if#1b\vbox \else \vcenter \fi\fi
  \bgroup \let\par\relax
  \let\@sharp##\let\protect\relax
  \@arrayskip\@preamble}
\makeatother

\newcommand{\newsection}[1]
{\vspace{4mm}
\pagebreak[3]
\addtocounter{section}{1}
\setcounter{equation}{0}
\setcounter{subsection}{0}
\begin{flushleft}
{\large\bf \thesection. #1}
\end{flushleft}
\nopagebreak
\medskip
\nopagebreak}

\newlength{\extraspace}
\newlength{\extraspaces}
\begin{document}

\renewcommand{\footnotesize}{\small}

\addtolength{\baselineskip}{.8mm}

\thispagestyle{empty}

\begin{flushright}
\baselineskip=12pt
DAMTP-98-36\\
DSF/16-98\\
OUTP-98-42P\\
hep-th/9806099\\
\hfill{  }\\ June 1998\\ Revised April 1999
\end{flushright}

\vskip 0.1in

\begin{center}

\baselineskip=12pt

{\Large\bf{String Geometry and the Noncommutative Torus}}\\[8mm]

{\bf Giovanni Landi}\footnote{On leave from Dipartimento di Scienze
Matematiche, Universit\`{a} di Trieste, P.le Europa 1, I-34127, Trieste, Italy.
Also INFN, Sezione di Napoli, Napoli, Italy.\\ E-mail: {\tt
landi@univ.trieste.it}}
\\[3mm]
{\it Department of Applied Mathematics and Theoretical Physics\\ University of
Cambridge, Silver Street, Cambridge CB3 9EW, U.K.}
\\[6mm]
{\bf Fedele Lizzi}\footnote{E-mail: {\tt fedele.lizzi@na.infn.it}}
\\[3mm]
{\it Dipartimento di Scienze Fisiche, Universit\`{a} di Napoli Federico II\\
and
INFN, Sezione di Napoli, Mostra d'Oltremare Pad.~20, 80125 Napoli, Italy}
\\[6mm]
{\bf Richard J.\ Szabo\footnote{Present address: The Niels Bohr
Institute, University of Copenhagen, Blegdamsvej 17, DK-2100 Copenhagen \O,
Denmark.\\ E-mail: {\tt szabo@nbi.dk}}}
\\[3mm]
{\it Department of Physics -- Theoretical Physics, University of Oxford\\ 1
Keble Road, Oxford OX1 3NP, U.K.}
\\[10mm]

{\sc Abstract}

\begin{center}
\begin{minipage}{16cm}

\baselineskip=12pt

We construct a new gauge theory on a pair of $d$-dimensional noncommutative
tori. The latter comes from an intimate relationship between the
noncommutative geometry associated with a lattice vertex operator algebra
$\alg$ and the noncommutative torus. We show that the tachyon
algebra of $\alg$ is naturally isomorphic to a class of twisted modules
representing quantum deformations of the algebra of functions on the torus.
We construct the corresponding real spectral triples and determine
their Morita equivalence classes using string duality arguments. These
constructions yield simple proofs of the $O(d,d;\zed)$ Morita equivalences
between $d$-dimensional noncommutative tori and give a natural physical
interpretation of them in terms of the target space duality group of
toroidally compactified string theory. We classify the automorphisms of the
twisted modules and construct the most general gauge theory which is
invariant under the automorphism group. We compute bosonic and fermionic
actions associated with these gauge theories and show that they are
explicitly duality-symmetric. The duality-invariant gauge theory is
manifestly covariant but contains highly non-local interactions. We show
that it also admits a new sort of particle-antiparticle duality which
enables the construction of instanton field configurations in any
dimension. The duality non-symmetric on-shell projection of the field
theory is shown to coincide with the standard non-abelian Yang-Mills gauge
theory minimally coupled to massive Dirac fermion fields.

\end{minipage}
\end{center}

\end{center}

\vfill
\newpage
\pagestyle{plain}
\setcounter{footnote}{0}
\setcounter{page}{1}
\stepcounter{subsection}

\newsection{Introduction}

The noncommutative torus \cite{Rieffel}--\cite{Rieffel1} is one of the basic
examples of a noncommutative geometry \cite{Book,giannibook} which captures
features of the difference between an ordinary manifold and a `noncommutative
space'. Recent interest in such geometries has occurred in the physics
literature in the context of their relation to $M$-Theory
\cite{cds}--\cite{Matrixmodule}. As shown in the seminal paper \cite{cds},
the most general solutions to the quotient conditions for toroidal
compactifications of Matrix Theory satisfy the algebraic relations of gauge
connections on noncommutative tori. This has led to, among other things, new
physical insights into the structure of the supergravity sector of $M$-Theory
by relating geometrical parameters of the noncommutative torus to physical
parameters of the Matrix Theory gauge theories. In this paper we shall discuss
the role played by the noncommutative torus in the short-distance structure of
spacetime. In particular we shall construct a new duality-symmetric gauge
theory on a pair of noncommutative tori.

One way to describe the noncommutative torus is to promote the ordinary
coordinates $x^i$, $i=1,\dots,d$, of a $d$-torus to non-commuting operators
$\hat x^i$ acting on an infinite-dimensional Hilbert space and obeying the
commutation relations
\beq
\left[\hat x^i,\hat x^j\right]=2\pi i\,\ell^{ij}
\label{nctoruscoords}\eeq
where $\ell^{ij}$ are real numbers. Defining the basic plane waves
\beq
U_i=\exp i\hat x^i
\label{planewaves}\eeq
it follows from the Baker-Campbell-Hausdorff formula that these operators
generate the algebra
\beq
U_iU_j=\e^{2\pi i\ell^{ij}}\,U_jU_i
\label{nctorusrels}\eeq
and these are the basic defining relations of the noncommutative torus,
when viewed
as an algebra of functions with generalized Fourier series expansions in terms
of the plane waves \eqn{planewaves}. In the limit $\ell^{ij}\to0$, one recovers
the ordinary coordinates and the plane waves \eqn{planewaves} become the usual
ones of the torus. The antisymmetric matrix $\ell^{ij}$ can therefore be
thought of as defining the Planck scale of a compactified spacetime. The
noncommutative torus then realizes old ideas of quantum gravity that at
distance scales below the Planck length the nature of spacetime geometry is
modified. At large distances the usual (commutative) spacetime is recovered.

On the other hand, a viable model for Planck scale physics is string
theory, and its various extensions to $D$-brane field theory
\cite{Dfield}--\cite{lms} and Matrix Theory \cite{Matrix}. In these latter
extensions, the short-distance noncommutativity of spacetime coordinates is
represented by viewing them as $N\times N$ matrices. The noncommutativity
of spacetime in this picture is the result of massless quantum excitations
yielding bound states of $D$-branes with broken supersymmetry
\cite{Dfield}, or alternatively of the quantum fluctuations in topology of
the worldsheets of the $D$-branes \cite{lms}. However, it is also possible
to view the spacetime described by ordinary string theory more directly as
a noncommutative geometry \cite{fg}--\cite{lslong}. The main idea in this
context is to substitute the (commutative) algebra of continuous
complex-valued functions on spacetime with the (noncommutative) vertex
operator algebra of the underlying conformal field theory. Ordinary
spacetime can be recovered by noticing that the algebra of continuous
functions can be thought of a subalgebra of the vertex operator algebra. The
fruitfulness of this approach is that the full vertex operator algebra is
naturally invariant under target space duality transformations of the string
theory. The simplest such mapping relates large and small radius circles to one
another and leads directly to a fundamental length scale, usually the
finite size of the string. Although the duality is a symmetry of the full
noncommutative string spacetime, different classical spacetimes are
identified under the transformation yielding a natural geometrical origin
for these quantum symmetries of compactified string theory. This point of
view therefore also describes physics at the Planck scale.

In this paper we will merge these two descriptions of the short-distance
structure of spacetime as described by noncommutative geometry. We
will show that a particular algebra obtained naturally from the tachyon
algebra, which in turn is obtained by projecting out the string oscillatory
modes, defines a twisted projective module over the noncommutative torus.
This algebra is, in this sense, the smallest quantum deformation of
ordinary, classical spacetime, and it represents the structure of spacetime
at the Planck length. Thus strings compactified on a torus have a geometry
which is already noncommutative at short distances.
%
%
The remainder of the full vertex operator algebra acts to yield the non-trivial
gauge transformations (including duality) of spacetime. This fact yields
yet another interpretation for the noncommutativity of Planck scale
spacetime. In string theory spacetime is a set of fields defined on a
surface, and at short distances the interactions of the strings (described
by the vertex operator algebra) causes the spacetime to become
noncommutative. In the case of toroidally compactified string theory,
spacetime at the the Planck scale is a noncommutative torus. The results of
this paper in this way merge the distinct noncommutative geometry
formalisms for strings by connecting them all to the geometry of the
noncommutative torus. At hand is therefore a unified setting for string
theory in terms of (target space) $D$-brane field theory, Matrix Theory
compactifications, and (worldsheet) vertex operator algebras.

Aside from these physically interesting consequences, we will show that the
modules we obtain from the vertex operator algebra also bear a number of
interesting mathematical characteristics. Most notably, the duality symmetries
of the vertex operator algebra lead to a simple proof of the Morita
equivalences of noncommutative tori with deformation parameters $\ell^{ij}$
which are related by the natural action of the discrete group $O(d,d;\zed)$ on
the space of real-valued antisymmetric $d\times d$ matrices. These Morita
equivalences have been established recently using more direct mathematical
constructions in \cite{Morita}. In \cite{schwarz} it was shown that Matrix
Theories compactified on Morita equivalent tori are physically equivalent to
one another, in that the BPS spectra of states are the same and the associated
field theories can be considered to be duals of each other. Here we shall find
a direct manifestation of this duality equivalence in terms of the basic
worldsheet theory itself. The relationship between vertex operator algebras,
duality and the noncommutative torus was originally pointed out in \cite{lsem}
and discussed further in \cite{lscastro}. In a sense, this relationship shows
that the duality properties of Morita equivalences in $M$-Theory are controlled
by the stringy sector of the dynamics. Morita equivalent noncommutative
tori have also been constructed in \cite{faddeev} under the name discrete
Heisenberg-Weyl groups and via the action of the modular group. There it
was also suggested that this is the base for the duality principles which
appear in string theory and conformal field theory.

We will also show that the fairly complete classification of the automorphism
group of the vertex operator algebra, given in \cite{lscastro}, can be used to
characterize the symmetries of the twisted modules that we find. This
immediately leads us to the construction of an action functional for this
particular noncommutative geometry which is naturally invariant under the
automorphism group. Generally, the action functional in noncommutative geometry
can be used to construct invariants of modules of the given algebra and it
presents a natural geometrical origin for many physical theories, such as the
standard model \cite{sm,connesreal} and superstring theory \cite{Cham}. The
spectral action principle of noncommutative geometry naturally couples
gravitational and particle interactions from a very simple geometric
perspective
\cite{connesauto,specaction}. In the following we shall construct both
fermionic and bosonic actions for the twisted modules which possess the same
properties as dictated by the spectral action principle. However, since we
shall be neither concerned with coupling to gravity nor in renormalization
effects, we shall use a somewhat simpler definition than that proposed in
\cite{specaction}. A consequence of the invariance of the action under
automorphisms of the algebra is that the action is explicitly
duality-symmetric. The construction of explicitly electric-magnetic duality
symmetric action functionals has been of particular interest over the years
\cite{zwan,deserteit} and they have had applications to the physics of
black holes
\cite{black} and of $D$-branes \cite{agan,cheung}. These actions are of special
interest now because of the deep relevance of duality symmetries to the
spacetime
structure of superstring theory within the unified framework of $M$-Theory.

Because the derivation of the duality-symmetric action involves quite a bit of
mathematics, it is worthwhile to summarize briefly the final result here. We
shall show that a general gauge theory on the twisted module leads naturally to
a target space Lagrangian of the form
\beq
{\cal L}=\left(F+{ }^\star F\right)_{ij}\left(F+{ }^\star
F\right)^{ij}-i\,\overline{\psi_*}\,\gamma^i\left(\partial_i+
i{\buildrel\leftrightarrow\over
{A_i}}\right)\,\psi-i\,\overline{\psi}\,\gamma^*_i\left(\partial_*^i+i
{\buildrel\leftrightarrow\over{A_*^i}}\right)\,\psi_*
\label{dualsymmaction}\eeq
where
\beq
F_{ij}=\partial_iA_j-\partial_jA_i+i\Bigl[A_i,A_j\Bigr]-g_{ik}\,g_{jl}
\left(\partial_*^kA_*^l-\partial_*^lA_*^k+i\left[A_*^k,A_*^l\right]\right)
\label{fieldstrengthsym}\eeq
and $i,j=1,\dots,d$. Here the field theory is defined on a Lorentzian spacetime
with coordinates
$(x^i,x_j^*)\in\real^d\times(\real^d)^*$, and $g_{ij}$ is the (flat) metric
of $\real^d$ while the metric on the total spacetime is $(g_{ij},-g_{ij})$. We
have defined $\partial_i=\partial/\partial x^i$ and
$\partial_*^j=\partial/\partial x^*_j$. The fields $A_i(x,x^*)$ and
$A_*^j(x,x^*)$ are a dual pair of gauge fields, $\psi(x,x^*)$ and
$\psi_*(x,x^*)$ are dual spinor fields, and $\gamma^i$ and $\gamma^*_i$ are
Dirac matrices on $\real^d$ and $(\real^d)^*$, respectively. The field strength
${ }^\star F_{ij}$ is a certain ``dual" to the field strength $F_{ij}$
with respect to the Lorentzian metric of the spacetime. The bars on the fermion
fields denote their ``adjoints" and the double arrow on the gauge potentials in
the fermionic part of the action denotes their left-right symmetric action on
the fermion and anti-fermion fields (in a sense which we shall define more
precisely in what follows). The commutators and actions of gauge potentials on
fermion fields are defined using the noncommutative tachyon algebra structure,
so that the action associated with \eqn{dualsymmaction} describes a certain
nonabelian gauge theory coupled to fermions in a nontrivial representation of a
gauge group (again this gauge group will be described more precisely in the
following).

The action corresponding to \eqn{dualsymmaction} is explicitly
invariant under the interchange of starred and un-starred
quantities. This symmetry incorporates the $T$-duality transformation
which inverts the metric $g_{ij}$, and it moreover contains a
particle-antiparticle duality transformation $F_{ij}\leftrightarrow{
}^\star F_{ij}$ that represents a certain topological instanton
symmetry of the field theory in any dimension $d$. However, by its
very construction, it is manifestly invariant under a much larger
symmetry group, including the gauge group, which we shall describe in
this paper. In this sense, we will see that the action functional
corresponding to \eqn{dualsymmaction} measures the amount of duality
symmetry as well as the strength of the string interactions present in
the given spacetime theory. Like the usual formulations of
electric-magnetic symmetric actions \cite{zwan} (see also
\cite{lsem}), \eqn{dualsymmaction} involves an $O(2,\real)$ doublet of
vector potentials $(A,A_*)$. The crucial difference between
\eqn{dualsymmaction} and the usual actions is that it is also {\it
manifestly covariant}, without the need of introducing auxiliary
fields \cite{deserteit}.  This general covariance follows from the
fact that the diffeomorphism symmetries of the spacetime are encoded
in the tachyon algebra as internal gauge symmetries, so that the gauge
invariance of the action automatically makes it covariant. As a
consequence of this feature, the on-shell condition for the field
strengths is different than those in the usual formulations. Here it
corresponds to a dimensional reduction
$\real^d\times(\real^d)^*\to\real^d$ in which the field theory becomes
ordinary Yang-Mills theory minimally coupled to massive Dirac
fermions. This reduction thus yields a geometrical origin for colour
degrees of freedom and fermion mass generation. The field theory
\eqn{dualsymmaction} can thereby be thought of as a first stringy
extension of many physical models, such as the standard model and
Matrix Theory. As a conventional field theory, however, the Lagrangian
\eqn{dualsymmaction} is highly non-local because it contains
infinitely many orders of derivative interactions through the
definition of the commutators (derived from the algebra
\eqn{nctorusrels}). This non-locality, and its origin as an
associative noncommutative product, reflects the nature of the string
interactions.

Thus, the natural action functional associated with the noncommutative
geometry of string theory not only yields an explicitly
duality-symmetric (non-local) field theory, but it also suggests a
sort of noncommutative Kaluza-Klein mechanism for the origin of
nonabelian gauge degrees of freedom and particle masses. The explicit
invariance of \eqn{dualsymmaction} under the duality group of the
spacetime thereby yields a physical interpretation of the mathematical
notion of Morita equivalence. This gauge theory on the noncommutative
torus is different from those discussed in the context of Matrix
Theory \cite{cds}--\cite{Matrixmodule} but it shares many of their
duality properties. The formalism of the present paper may thus be
considered as a step towards the formulation of Matrix Theory in terms
of the framework of spectral triples in noncommutative geometry
\cite{howu}. It is a remarkable feature that such target space
dynamics can be induced so naturally at the level of a worldsheet
formalism.

The structure of the remainder of this paper is as follows. All ideas and
results of noncommutative geometry which we use are briefly explained
throughout the paper. In section 2 we will briefly define the vertex operator
algebra associated with toroidally compactified string theory. In section 3 we
study the tachyon algebra of the lattice vertex operator algebra and show
that it defines a particular twisted module over the noncommutative torus. In
section 4 we introduce a set of spectral data appropriate to the noncommutative
geometry of string theory. In section 5 we exploit the duality symmetries of
this noncommutative geometry to study some basic properties of the twisted
module, including its Morita equivalence classes and its group of
automorphisms. In section 6 the most general gauge theory on the module is
constructed, and in section 7 that gauge theory is used to derive the
duality-symmetric action functional. In section 8 we describe some heuristic,
physical aspects of these twisted modules along the lines described in this
section. For completeness, an appendix at the end of the paper gives a brief
overview of the definition and relevant mathematical significance of Morita
equivalence in noncommutative geometry.

\newsection{Lattice Vertex Operator Algebras}\label{se:lvoa}

In this section we will briefly review, mainly to introduce notation, the
definition of a lattice vertex operator algebra (see \cite{lslong,flm} and
references therein for more details). Let $L$ be a free infinite
discrete abelian group of rank $d$ with $\zed$-bilinear form
$\langle\,\cdot\,,\,\cdot\,\rangle_L:L\times L\to\real^+$ which is symmetric
and nondegenerate. Given a basis $\{e_i\}_{i=1}^d$ of $L$, the symmetric
nondegenerate tensor
\beq
g_{ij}=\left\langle e_i,e_j\right\rangle_L
\label{metricdef}\eeq
defines a Euclidean metric on the flat $d$-dimensional torus
$T_d\equiv\real^d/2\pi L$. We can extend the inner product
$\langle\,\cdot\,,\,\cdot\,\rangle_L$ to the complexification
$L^c=L\otimes_\zeds\complex$ by $\complex$-linearity. The dual lattice to $L$
is then
\beq
L^*=\left\{p\in L^c~|~\langle p,w\rangle_L\in\zed~~\forall w\in L\right\}
\label{duallatticedef}\eeq
which is also a Euclidean lattice of rank $d$ with bilinear form $g^{ij}$
inverse to \eqn{metricdef} that defines a metric on the dual torus
$T_d^*\equiv\real^d/2\pi L^*$.

Given the lattice $L$ and its dual \eqn{duallatticedef}, we can form the free
abelian group
\beq
\Lambda=L^*\oplus L
\label{Lambdadef}\eeq
If $\{e^i\}_{i=1}^d$ is a basis of $L^*$ dual to a basis $\{e_i\}_{i=1}^d$ of
$L$, then the chiral basis of $\Lambda$ is $\{e^i_\pm\}_{i=1}^d$, where
\beq
e^i_\pm=\mbox{$\frac1{\sqrt2}$}\left(e^i\pm g^{ij}e_j\right)
\label{chiralbasis}\eeq
with an implicit sum over repeated indices always understood. Given $p\in L^*$
and $w\in L$, we write the corresponding elements of $\Lambda$ with respect to
the basis \eqn{chiralbasis} as $p^\pm$ with components
\beq
p^\pm_i=\mbox{$\frac1{\sqrt2}$}\Bigl(p_i\pm\langle e_i,w\rangle_L\Bigr)
\label{chiralmom}\eeq
Then with $q_i^\pm=\frac1{\sqrt2}(q_i\pm\langle e_i,v\rangle_L)$, we can
define
a $\zed$-bilinear form
$\langle\,\cdot\,,\,\cdot\,\rangle_\Lambda:\Lambda\times\Lambda\to\zed$ by
\beq
\left\langle p,q\right\rangle_\Lambda\equiv
p_i^+g^{ij}q_j^+-p_i^-g^{ij}q_j^-=\langle p,v\rangle_L+\langle q,w\rangle_L
\label{Lambdametricdef}\eeq
and this makes $\Lambda$ an integral even self-dual Lorentzian lattice of
rank $2d$ and signature $(d,d)$ which is called the Narain lattice. Note
that, in
the chiral basis \eqn{chiralbasis}, the corresponding metric tensor is
\beq
\eta_{\alpha\beta}^{ij}\equiv\left\langle
e_\alpha^i,e_\beta^j\right\rangle_\Lambda=\left\{\new{\begin{array}{rrl}\pm
\,g^{ij}~~&,&~~\alpha=\beta=\pm\\0~~&,&~~\alpha\neq\beta\end{array}}\right.
\label{chiralmetric}\eeq

The commutator map
\beq
c_\Lambda(p^+,p^-;q^+,q^-)=\e^{i\pi\langle q,w\rangle_L}
\label{cocycledef}\eeq
on $\Lambda^c\times\Lambda^c\to\zed_2$ is a two-cocycle of the group algebra
$\complex[\Lambda]$ generated by the vector space $\Lambda^c$,
the complexification of $\Lambda$. This two-cocycle corresponds
to a central extension $\widetilde{\Lambda^c}$ of $\Lambda^c$,
\beq
1~\to~\zed_2~\to~\widetilde{\Lambda^c}~\to~\Lambda^c~\to~1
\label{centralext}\eeq
where $\widetilde{\Lambda^c}=\zed_2\times\Lambda^c$ as a set and the
multiplication is given by
\beq
\left(\rho\,;\,q^+,q^-\right)\cdot\left(\sigma\,;\,r^+,r^-\right)=\left
(c_\Lambda(q^+,q^-;r^+,r^-)\rho\sigma\,;\,q^++r^+,q^-+r^-\right)
\label{latticemult}\eeq
for $\rho,\sigma\in{\bb Z}_2$ and $(q^+,q^-),(r^+,r^-)\in\Lambda^c$. This can
be used to define the twisted group algebra $\complex\{\Lambda\}$ associated
with the double cover $\widetilde{\Lambda^c}$ of $\Lambda^c$. A realization of
$\complex\{\Lambda\}$ in terms of closed string modes is given as follows.
Viewing $L^c$ and $(L^*)^c$ as abelian Lie algebras of dimension $d$, we can
consider the corresponding affine Lie algebras $\widehat{L^c}$ and
$\widehat{(L^*)^c}$, and also the affinization
$\widehat{\Lambda^c}=\widehat{L^c}\oplus\widehat{(L^*)^c}$. In the basis
\eqn{chiralbasis}, we then have
\beq
\widehat{\Lambda^c}\cong\widehat{u(1)_+^d}\oplus\widehat{u(1)_-^d}
\label{affineLambda}\eeq
where the generators $\alpha_n^{(\pm)i}$, $n\in\zed$, of
$\widehat{u(1)_\pm^d}$
satisfy the Heisenberg algebra
\beq
\left[\alpha_n^{(\pm)i},\alpha_m^{(\pm)j}\right]=n\,g^{ij}\,\delta_{n+m,0}
\label{heisalg}\eeq

The basic operators of interest to us are the chiral Fubini-Veneziano fields
\beq
X_\pm^i(z_\pm)=x_\pm^i+ig^{ij}p_j^\pm\log
z_\pm+\sum_{n\neq0}\frac1{in}\,\alpha_n^{(\pm)i}z_\pm^{-n}
\label{FVfieldsdef}\eeq
and the chiral Heisenberg fields
\beq
\alpha_\pm^i(z_\pm)=-i\partial_{z_\pm}X_\pm^i(z_\pm)=
\sum_{n=-\infty}^\infty
\alpha_n^{(\pm)i}z_\pm^{-n-1}~~~~~~;~~~~~~\alpha_0^{(\pm)i}\equiv
g^{ij}p_j^\pm
\label{heisfieldsdef}\eeq
where $z_\pm\in\complex\cup\{\infty\}$. Classically, the fields
\eqn{FVfieldsdef} are maps from the Riemann sphere into the torus $T_d$, with
$x_\pm\in T_d$. When they are interpreted as classical string embedding
functions from a cylindrical
worldsheet into a toroidal target space, the first two terms in
\eqn{FVfieldsdef} represent the center of mass (zero mode) motion of a
closed string while the Laurent series represents its oscillatory
(vibrational) modes. The $w^i$ in
\eqn{chiralmom} represent the winding modes of the string about each of the
cycles of $T_d$, while the fields
\eqn{heisfieldsdef} are the conserved currents which generate infinitesimal
reparametrizations of the torus.

Upon canonical quantization, the oscillatory modes satisfy \eqn{heisalg} and
the zero modes form a canonically conjugate pair,
\beq
\left[x_\alpha^i,p_j^\beta\right]=i\,\delta_\alpha^\beta\,\delta_i^j
\label{zeromodequant}\eeq
where $\alpha,\beta=\pm$. Then the multiplication operators
\beq
\varepsilon_{q^+q^-}(p^+,p^-)\equiv\e^{-iq_i^+x_+^i-iq_i^-x_-^i}~
c_\Lambda(p^+,p^-;q^+,q^-)
\label{twistedgens}\eeq
generate the twisted group algebra $\complex\{\Lambda\}$. The fields $X_\pm$
now become quantum operators which act formally on the infinite-dimensional
Hilbert space
\beq
{\bf h}=L^2\left(T_d\times
T_d^*\,,\,\mbox{$\prod_{i=1}^d\frac{dx^i\,dx^*_i}{(2\pi)^2}$}\right)
\otimes{\cal F}^+\otimes{\cal F}^-
\label{hilbertdef}\eeq
where $x^i\equiv\frac1{\sqrt2}(x_+^i+x_-^i)$ and
$x_i^*\equiv\frac1{\sqrt2}\,g_{ij}(x_+^j-x_-^j)$ define, respectively, local
coordinates on the torus $T_d$ and on the dual torus $T_d^*$. The $L^2$ space
in \eqn{hilbertdef} is generated by the canonical pairs \eqn{zeromodequant} of
zero modes and is subject to the various natural isomorphisms
\beq\new{\begin{array}{lll}
L^2\left(T_d\times
T_d^*\,,\,\mbox{$\prod_{i=1}^d\frac{dx^i\,dx^*_i}{(2\pi)^2}$}\right)&\cong&
L^2\left(T_d\,,\,\mbox{$\prod_{i=1}^d\frac{dx^i}{2\pi}$}\right)\otimes_
\complexs
L^2\left(T_d^*\,,\,\mbox{$\prod_{i=1}^d\frac{dx^*_i}{2\pi}$}\right)\\&\cong&
\bigoplus_{w\in
L}L^2\left(T_d\,,\,\mbox{$\prod_{i=1}^d\frac{dx^i}{2\pi}$}\right)\\
&\cong&\bigoplus_{p\in L^*}L^2\left(T_d^*\,,\,
\mbox{$\prod_{i=1}^d\frac{dx^*_i}{2\pi}$}\right)\end{array}}
\label{hilbertisos}\eeq
The Hilbert space \eqn{hilbertdef} is thus a module for the group algebras
$\complex[L]$ and $\complex[L^*]$, and also for $\complex\{\Lambda\}$. The
dense subspace $C^\infty(T_d\times T_d^*)\subset L^2(T_d\times
T_d^*\,,\,\prod_{i=1}^d\frac{dx^i\,dx^*_i}{(2\pi)^2})$ of smooth complex-valued
functions on $T_d\times T_d^*$ is a unital $*$-algebra which is spanned by the
eigenstates $|q,v\rangle=|q^+,q^-\rangle=\e^{-iq_ix^i-iv^ix_i^*}$ of the
operators $-i\frac\partial{\partial x^i}$ and $-i\frac\partial{\partial x^*_i}$
on $T_d\times T_d^*$, where $q\in L^*$ and $v\in L$. The spaces ${\cal F}^\pm$
are bosonic Fock spaces generated by the oscillatory modes $\alpha_n^{(\pm)i}$
which act as annihilation operators for $n>0$ and as creation operators for
$n<0$ on some vacuum states $|0\rangle_\pm$.

The basic single-valued quantum fields which act on the Hilbert space
\eqn{hilbertdef} are the chiral tachyon operators
\beq\new{
V_{q^\pm}(z_\pm)=\NO\e^{-iq_i^\pm X_\pm^i(z_\pm)}\NO}
\label{chiraltachyonops}\eeq
where $(q^+,q^-)\in\Lambda$ and $\NO\cdot\NO$ denotes the Wick normal ordering
defined by reordering the operators (if necessary) so that all
$\alpha_n^{(\pm)i}$, $n<0$, and $x_\pm^i$ occur to the left of all
$\alpha_n^{(\pm)i}$, $n>0$, and $p^\pm_i$. The twisted dual
$\widetilde{{\bf h}^*}$ of the Hilbert space \eqn{hilbertdef} is spanned by
operators of the form
\beq
\Psi=\varepsilon_{q^+q^-}(p^+,p^-)\otimes\prod_k\alpha_{-n_k}^{(+)i_k}\otimes
\prod_l\alpha_{-m_l}^{(-)j_l}
\label{Psidual}\eeq
To \eqn{Psidual} we associate the vertex operator
\beq\new{
V(\Psi;z_+,z_-)=\NO
V_{q^+q^-}(z_+,z_-)~\prod_k\frac1{(n_k-1)!}\,\partial_{z_+}^{n_k-1}
\alpha_+^{i_k}(z_+)~\prod_l\frac1{(m_l-1)!}\,\partial_{z_-}^{m_l-1}
\alpha_-^{j_l}(z_-)\NO}
\label{vertexopdef}\eeq
where
\beq\new{\begin{array}{lll}
V_{q^+q^-}(z_+,z_-)&\equiv&
V\left(\varepsilon_{q^+q^-}(p^+,p^-)\otimes\id;z_+,z_-\right)\\&=&
c_\Lambda(p^+,p^-;q^+,q^-)\,\NO
V_{q^+}(z_+)\,V_{q^-}(z_-)\NO\\&=&\e^{i\pi\langle
q,w\rangle_L}\,\NO\e^{-iq_i^+X_+^i(z_+)-iq_i^-X_-^i(z_-)}\NO\end{array}}
\label{tachyondef}\eeq
are the basic tachyon vertex operators. This gives a well-defined linear map
$\Psi\mapsto V(\Psi;z_+,z_-)$ on $
\widetilde{{\bf h}^*}\to({\rm End}\,{\bf h})[z_+^{\pm1},z_-^{\pm1}]$,
the latter being the space of endomorphism-valued Laurent polynomials in
the variables $z_\pm$. This mapping is known as the operator-state
correspondence.

With an appropriate regularization (see the next section), the vertex operators
\eqn{vertexopdef} yield well-defined and densely-defined operators acting on
$\bf h$. They generate a noncommutative unital $*$-algebra $\alg$ with the
usual Hermitian conjugation and operator norm (defined on an appropriate dense
domain of bounded operators in $\alg$). The various algebraic properties of
$\alg$ can be found in \cite{lslong,flm}, for example. It has the formal
mathematical structure of a vertex operator algebra. In this paper we shall not
discuss these generic properties of $\alg$, but will instead analyse in detail
a particular `subalgebra' of it which possesses some remarkable properties.

It is important to note that $\alg\cong\widetilde{{\cal E}_\Lambda}$ is the
$\zed_2$-twist of the algebra ${\cal E}_\Lambda={\cal
E}_+\otimes_\complexs{\cal E}_-$, where ${\cal E}_\pm$ are the chiral
algebras
generated by the operators \eqn{chiraltachyonops}. The lattice $L^*$ itself
yields the structure of a vertex operator algebra ${\cal E}_{L^*}$ without any
reference to its dual lattice or the Narain lattice with its chiral sectors.
Indeed, given a basis $\{e^i\}_{i=1}^d$ of $L^*$ and two-cocycles
$c_L(p,q)=\e^{i\pi\langle q,p\rangle_{L^*}}$ corresponding to an $S^1$ covering
group of the dual lattice, the operators
\beq\new{
\widetilde{V}_q(z)=\e^{i\pi\langle q,p\rangle_{L^*}}\,\NO\e^{-iq_iX^i(z)}\NO}
\label{Ltachyons}\eeq
generate a vertex operator algebra, acting on the Hilbert space
$L^2(T_d^*\,,\,\prod_{i=1}^d\frac{dx^*_i}{2\pi})\otimes{\cal F}$, in an
analogous way that the operators \eqn{tachyondef} do \cite{flm}. Similarly one
defines a vertex operator algebra ${\cal E}_L$ associated with the lattice $L$.
Modulo the twisting factors, the algebras $\alg$ and ${\cal
E}_{L^*}\otimes_\complexs{\cal E}_L$ are isomorphic since the latter
representation comes from changing basis on $\Lambda$ from the chiral one
\eqn{chiralbasis} to the canonical one $\{e^i\}_{i=1}^d\oplus\{e_j\}_{j=1}^d$.
Physically, the difference between working with the full chirally symmetric
algebra $\alg$ and only the chiral ones or ${\cal E}_{L^*}$, ${\cal E}_L$ is
that the former represents the algebra of observables of closed strings while
the latter ones are each associated with open strings. Furthermore, the algebra
$\alg$ is that which encodes the duality symmetries of spacetime and maps the
various open string algebras among each other via duality transformations. Much
of what is said in the following will therefore not only apply as symmetries of
a closed string theory, but also as mappings among various open string
theories. It is in this way that these results are applicable to $D$-brane
physics and $M$-Theory.

\newsection{Tachyon Algebras and Twisted Modules over the Noncommutative
Torus}\label{se:tamnct}

The operator product algebra of the chiral tachyon operators
\eqn{chiraltachyonops} can be evaluated using standard normal ordering
properties to give \cite{gsw}
\beq\new{
V_{q^\pm}(z_\pm)\,V_{r^\pm}(z'_\pm)=
\left(z_\pm-z'_\pm\right)^{q_i^\pm
g^{ij}r_j^\pm}\,\NO V_{q^\pm}(z_\pm)\,V_{r^\pm}(z'_\pm)\NO}
\label{vertexopalg1}\eeq
which leads to the operator product
\beq\new{\begin{array}{lll}
V_{q^+q^-}(z_+,z_-)\,V_{r^+r^-}(z'_+,z'_-)&=&\left(z_+-z'_+\right)^
{q_i^+g^{ij}r_j^+}\left(z_--z'_-\right)^{q_i^-g^{ij}r_j^-}\\& &~~~~\times
\NO V_{q^+q^-}(z_+,z_-)\,V_{r^+r^-}(z'_+,z'_-)\NO\end{array}}
\label{vertexopalg2}\eeq
of the full vertex operators \eqn{tachyondef}.
They are the defining characteristics of the
vertex operator algebra which in the mathematics literature are
collectively referred to as the Jacobi identity of $\alg$
\cite{lslong,flm}. The product of a tachyon operator with its Hermitian
conjugate is given by
\bea
V_{q^\pm}(z_\pm)\,V_{q^\pm}(z'_\pm)^\dagger&=&\left({z_\pm\over
z_\pm'}\right)^{p_i^\pm
g^{ij}q_j^\pm} \left(1-{z_\pm'\over z_\pm}\right)^{q_i^\pm g^{ij} q_j^\pm}
\prod_{n>0}\e^{q_i^\pm{\alpha_{-n}^{(\pm)i}\over n}(z_\pm^n-{z_\pm'}^n)}\nn\\
& &\times\prod_{n>0}\e^{q_i^\pm
{\alpha_{n}^{(\pm)i}\over n}(z_\pm^{-n}-{z_\pm'}^{-n})}
\label{vvdag}\eea

The tachyon vertex operators thus form a $*$-algebra which,
according to the operator-state correspondence of the previous section, is
associated with the subspace ${\bf h}_0\subset\bf h$ defined by
\beq
{\bf h}_0~\equiv~\bigotimes_{n>0}\,
\bigotimes_{i=1}^d\left(\ker\alpha_n^{(+)i}\otimes\ker
\alpha_n^{(-)i}\right)~\cong~L^2\left(T_d\times
T_d^*\,,\,\mbox{$\prod_{i=1}^d\frac{dx^i\,dx^*_i}{(2\pi)^2}$}\right)
\label{H0def}\eeq
Algebraically, ${\bf h}_0$ is the subspace of highest weight vectors for the
representation of $\widehat{u(1)_+^d}\oplus\widehat{u(1)_-^d}$ on the $L^2$
space, and it is spanned by the vectors $|q^+;q^-\rangle$. The irreducible
(highest weight) representations of this current algebra are labelled by the
$u(1)_\pm^d$ charges $q_i^\pm$, and ${\bf h}_0$ carries a representation of
$u(1)_+^d\oplus u(1)_-^d$ given by the actions of $\alpha_0^{(\pm)i}$. Then,
\beq
{\bf h}=\left(\bigoplus_{(q^+,q^-)\in\Lambda}{\cal
V}_{q^+q^-}\right)\otimes{\cal F}^+\otimes{\cal F}^-
\label{Hirrep}\eeq
where $q_i^\pm=\frac1{\sqrt2}(q_i\pm\langle e_i,v\rangle_L)$ and ${\cal
V}_{q^+q^-}\cong{\cal V}_q\otimes{\cal V}_v$ are the irreducible
$u(1)_+^d\oplus u(1)_-^d$ modules for the representation of the Kac-Moody
algebra on $L^2(T_d\times
T_d^*\,,\,\prod_{i=1}^d\frac{dx^i\,dx^*_i}{(2\pi)^2})$. In this sense, the
tachyon operators generate the full vertex operator algebra $\alg$, and the
algebraic relations between any set of vertex operators \eqn{vertexopdef} can
be deduced from the operator product formula \eqn{vertexopalg2} \cite{lslong}.
In the following we will therefore focus on the tachyon sector of the vertex
operator algebra.

If ${\cal P}_0:{\bf h}\to{\bf h}_0$ is the orthogonal projection onto the
subspace \eqn{H0def}, then the low energy tachyon algebra is
defined to be
\beq
\alg_0={\cal P}_0\,\alg\,{\cal P}_0
\label{tachyonalgdef}\eeq
If we consider only the zero mode part of the tachyon operators (by
projecting out the oscillatory modes)
then the corresponding algebra generators are
given by
\beq
{\cal P}_0\,V_{q_+q_-}(z_+,z_-)\,{\cal P}_0=K(z_+,z_-)~\e^{i\pi\langle
q,w\rangle_L}~\e^{-iq_ix^i-iv^ix_i^*}=K(z_+,z_-)\,\varepsilon_{q^+q^-}(p^+,p^-)
\label{subalggens}\eeq
with $K(z_+,z_-)$ a worldsheet dependent normalization equal to unity at
$z_+=z_-=1$.
The generators of $\alg_0$ coincide with the generators \eqn{twistedgens}
of $\complex\{\Lambda\}$.
{}From the multiplication property
\beq
\varepsilon_{q^+q^-}(p^+,p^-)\,\varepsilon_{r^+r^-}(p^+,p^-)
=c_\Lambda(q^+,q^-;r^+,r^-)~\varepsilon_{(q^++r^+)(q^-+r^-)}(p^+,p^-)
\label{twistedrels}\eeq
we find the clock algebra
\beq
\varepsilon_{q^+q^-}(p^+,p^-)\,\varepsilon_{r^+r^-}(p^+,p^-)=\e^{i\pi
\gamma_\Lambda(q^+,q^-;r^+,r^-)}~\varepsilon_{r^+r^-}(p^+,p^-)\,
\varepsilon_{q^+q^-}(p^+,p^-)
\label{epclockalg}\eeq
where
\beq
\gamma_\Lambda(q^+,q^-;r^+,r^-)=\langle r,v\rangle_L-\langle q,w\rangle_L
\label{gammacocycledef}\eeq
is a two-cocycle of $\Lambda^c$.
This sector represents the extreme low energy of the string theory, in which
the oscillator modes
are all turned off. In this limit the geometry is a ``commutative'' one and
the physical theory is that of a point
particle theory. All stringy effects have been eliminated. In the following
we will attempt to go beyond this limit.

The expressions \eqn{vertexopalg1}--\eqn{vvdag} are singular at
$z_\pm=z'_\pm$ and are therefore to be understood only as formal
relationships valid away from coinciding positions of the operators on
$\complex\cup\{\infty\}$. Moreover the vertex operators, seen as operators
on the Hilbert space, are in general not bounded, and to define a
$C^*$-algebra one usually `smears' them\footnote{For issues related to the
boundedness of vertex operators see \cite{ConstScharf}.}.
Nevertheless the ultraviolet divergence related to the product of the
operators at coinciding points can be cured by introducing a cutoff on the
worldsheet, a common practice in conformal field theory. We will implement this
cutoff by considering a ``truncated'' algebra obtained by considering only a
finite number of oscillators. With this device we solve the unboundedness
problem
as well. We define
\beq
V_{q^\pm}(z_\pm)=\prod_{n\geq 0} W_n^\pm(z_\pm)
\eeq
where $W_0^\pm$ contains the zero modes $x_\pm$ and $p^\pm$, while the
$W_n^\pm$'s for $n\neq0$ involve only the $n^{\rm th}$ oscillator modes
$\alpha_{n}^{(\pm)i}$ and $\alpha_{-n}^{(\pm)i}$. The truncated vertex operator
is then defined as
\beq
V_{q^\pm}^{N}(z_\pm)={\cal N}_N^\pm\prod_{n=0}^N W_n^\pm(z_\pm)
\label{cutvertexop}\eeq
The quantity ${\cal N}_N^\pm$ is a normalization constant which we choose to be
${\cal N}_N^\pm=\prod_{n=1}^N\e^{-q_i^\pm g^{ij}q_j^\pm/n}$. With this
normalization the operators \eqn{cutvertexop} are unitary. There is now no
problem in multiplying operators at coinciding points, but the relation
\eqn{vertexopalg1} will change and be valid only in the limit $N\to\infty$.

The operator product formula \eqn{vertexopalg1} (and its modification in
the finite $N$ case) imply a cocycle relation among the operators
$V_{e^i_\pm}^N$. Interchanging the order of the two operators and using
standard tricks of operator product expansions \cite{gsw}), it follows that the
truncated algebra is
generated by the elements $V_{e^i_\pm}(z_i)$ subject to the relations
\bea
V^N_{e^i_\pm}({z_\pm}_i)\,V^N_{e^j_\mp}({z_\mp}_j)&=&V^N_{e^j_\mp}({z_\mp}_j)
\,V^N_{e^i_\pm}({z_\pm}_i)\label{VVpm}\\
V^N_{e^i_\pm}({z_\pm}_i)\,{V^N_{e^i_\pm}({z_\pm}_i)}^\dagger&=&
{V^N_{e^i_\pm}({z_\pm}_i)}^\dagger
\,V^N_{e^i_\pm}({z_\pm}_i)~=~\id\label{VV*}\\
V^N_{e^i_\pm}({z_\pm}_i)\,
V^N_{e^j_\pm}({z_\pm}_j)&=&\e^{2\pi
i{\omega_N}_\pm^{ij}}~V^N_{e^j_\pm}({z_\pm}_j)\,V^N_{e^i_\pm}({z_\pm}_i)
{}~~~~,~~~~i\neq j
\label{cocyclelocrel}\eea
where $\dagger$ is the $*$-involution on $\alg$, the
${z_\pm}^i$ are distinct points, and
\beq
{\omega_N}_\pm^{ij}=\pm\,g^{ij}\left[\log\left({z^\pm_i\over z^\pm_j}\right)-
\sum_{n=1}^N{1\over n}\left(\left({z^\pm_j\over z^\pm_i}\right)^n
-\left({z^\pm_i\over z^\pm_j}\right)^n\right)\right]\ \ . \label{omegan}
\eeq

The mutually commuting pair of (identical) algebras
\eqn{VV*},\eqn{cocyclelocrel} consists of two copies $\alg^{(\omega_N)}_+ \cong
\alg^{(\omega_N)}_- \cong \alg^{(\omega_N)}$ of the algebra of the
noncommutative $d$-torus $T_{\omega_N}^d$. Choosing an appropriate closure, the
algebra $\alg^{(\omega_N)}$ can be identified with the abstract unital
$*$-algebra
generated by elements $U_i~,i=1,\dots,d$, subject to the cocycle relations
\eqn{VV*},\eqn{cocyclelocrel},
\bea
U_i \,U_i^\dagger&=& U_i^\dagger \,U_i~=~\id \\ U_i \, U_j &=&\e^{2\pi
i\omega_N^{ij}}~U_j\,U_i {}~~~~,~~~~i\neq j
\label{cocycletorus}\eea
The presentation of a generic `smooth' element of $\alg^{(\omega_N)}$ in
terms of the $U_i$ is
\beq
f=\sum_{p\in L^*}f_p~U_1^{p_1}U_2^{p_2}\cdots U_d^{p_d}
{}~~~~~~,~~~~~~f_p\in{\cal S}(L^*)
\label{Fourierexp}\eeq
where ${\cal S}(L^*)$ is a Schwartz space of rapidly decreasing sequences.
When $\omega_N=0$, one can identify the operator $U_i$ with the
multiplication by $U_i=\e^{-ix^i}$ on the ordinary torus $T_d$, so that
$\alg^{(0)} \cong C^\infty(T_d)$, the algebra of smooth complex-valued
functions on $T_d$. In this case the expansion \eqn{Fourierexp} reduces to
the usual Fourier series expansion $f(x)=\sum_{p\in L^*}f_p~\e^{-ip_ix^i}$.

For a general non-vanishing anti-symmetric bilinear form $\omega_N$, we
shall think of the algebra $\alg^{(\omega_N)}$ as a quantum deformation of
the algebra $C^\infty(T_d)$. While the $*$-involution is the usual
complex-conjugation $f^\dagger(x)=\overline{f(x)}$, given $f,g\in
C^\infty(T_d)$ their deformed product is defined to be
\beq (f\star_{\omega_N}
g)(x)=\exp\left(i\pi\omega_N^{ij}\,\mbox{$\frac\partial{\partial
x^i}\frac\partial{\partial x'^j}$}\right)f(x)g(x')\Bigm|_{x'=x}
\label{deformedprod}\eeq
Furthermore, the unique normalized trace $\tau
:\alg^{(\omega_N)}\to\complex$ is represented by the classical average
\cite{debievre}
\beq
\tau(f)= \int_{T_d}\,\prod_{i=1}^d\frac{dx^i}{2\pi}~f(x)
\label{nctorustrace}\eeq
which is equivalent to projecting onto the zero mode $f_0$ in the Fourier
series expansion \eqn{Fourierexp}, $\tau(f)=f_0$. This trace will be used later
on to construct a duality-invariant gauge theory.

Once we have established the connection (at finite $N$) between the Vertex
Operator Algebra and the Noncommutative Torus we can then take the limit
$N\to\infty$. Then, the cocycle ${\omega_N}_\pm$ defined in
\eqn{omegan} converges to
\beq
\omega_\pm^{ij}=\pm\,g^{ij}~{\rm sgn}({\rm arg}\,z_i^\pm-{\rm arg}\,z_j^\pm)
{}~~~~~~,~~~~~~i\neq j~.
\label{cocycleloc}\eeq
With $\omega_\pm^{ii}\equiv0$ we obtain two-cocyles. We have chosen the
branch on $\complex\cup\{\infty\}$ of the logarithm function for which the
imaginary part of $\log(-1)$ lies in the interval $[-i\pi,+i\pi]$. The
cocycles ${\omega}_\pm$ depend only on the relative orientations of the
phases of the given tachyon operators. If we order these phases so that
$\pm\arg z_1^\pm>\pm\arg z_2^\pm>\dots>\pm\arg z_d^\pm$, then
${\omega}_\pm^{ij}=\omega^{ij}\equiv{\rm sgn}(j-i)\,g^{ij}$ for $i\neq j$.
This choice is unique up to a permutation in $S_{2d}$ of the coordinate
directions of $T_d\times T_d^*$. Thus we obtain the noncommutative torus
characterized by $\omega$ defined in \eqn{cocycleloc}.

We will indicate
with $\alg^{(\omega)}$ the algebra obtained in this fashion and will
consider this
algebra to act on the tachyon Hilbert space, which is motivated by the fact
that
at up to Planckian energies only the tachyonic states are excited. Of course at
higher energies also the oscillatory (Fock space) modes of the Hilbert
space will
have to be considered.
The algebra $\alg^{(\omega)}$ that we are considering can also be seen as a
deformation of $\alg_0$ defined in \eqn{tachyonalgdef}, in which the
product of the
elements has to be made in the full algebra, and then the result is to be
projected on the tachyon Hilbert space. Namely, given $V_0={\cal P}_0V{\cal
P}_0,W_0={\cal P}_0W{\cal P}_0\in\alg_0$, the operator product expansion with
the oscillators, gives non-trivial relations among the tachyons which we
identify with the usual relations of the noncommutative torus so that we can
define a deformed product by
\beq
V_0*W_0={\cal P}_0\,(VW)\,{\cal P}_0\ \ .
\eeq
This is a very natural way to deform the tachyonic algebra ${\cal A}_0$
which takes the presence of oscillator modes into account, the projection
operators being positioned in such a way to ensure that all oscillatory
contribute to the product. The $*$-algebra $\alg^{(\omega)}$ therefore
defines a
module for the noncommutative torus which possesses some remarkable properties
that distinguish it from the usual modules for $T_\omega^d$. Although it is
related to the projective regular representation of the twisted group algebra
$\complex\{\Lambda\}$ of the Narain lattice, this is not the algebraic feature
which determines the cocycle relation \eqn{cocyclelocrel}. The clock algebra
\eqn{epclockalg} is very different from the algebra defined by
\eqn{cocyclelocrel}
which arises from the operator product \eqn{vertexopalg1}.
The latter algebra is actually associated
with the twisted chiral operators $V_{e^\pm_i}$, although their
chiral-antichiral products do not yield the
$\zed_2$-twist of the tachyon vertex operators \eqn{subalggens}.
 Thus the product of the two algebras in
\eqn{VVpm}--\eqn{cocyclelocrel} have deformation matrix associated with the
bilinear form \eqn{chiralmetric}. However, the minus sign which appears in
the antichiral sector is irrelevant since $T_{-\omega}^d\cong T_\omega^d$
by a relabeling of the generators $U_i$.

The algebra $\alg^{(\omega)}$
thus defines a $2d$-dimensional $\zed_2$-twisted module
$\widetilde{\cal T}_\omega^d$ over the noncommutative torus, where ${\cal
T}_\omega^d$ carries a double representation $T^{(+)d}_\omega\times
T^{(-)d}_\omega$ of $T_\omega^d$. The non-trivial $\zed_2$-twist of this module
is important from the point of view of the noncommutative geometry of the
string spacetime. Note that its algebra is defined by computing the operator
product relations in \eqn{VVpm}--\eqn{cocyclelocrel} in the full vertex
operator algebra $\alg$, and then {\it afterwards} projecting onto the tachyon
sector. Otherwise we arrive at the (trivial) clock algebra, so that
within the definition of $\widetilde{\cal T}_\omega^d$ there is a particular
ordering that must be carefully taken into account. This module is therefore
quite different from the usual projective modules over the noncommutative torus
\cite{ncmodules,Matrixmodule}, and we shall exploit this fact dramatically in
what follows. Note that if we described $\alg$ using instead the canonical
basis $\{e^i\}_{i=1}^d\oplus\{e_j\}_{j=1}^d$ of the Narain lattice $\Lambda$,
i.e. taking as generators for $\alg$ twisted products of operators of the form
\eqn{Ltachyons} and their duals, then we would have arrived at a twisted
representation of $T_\omega^d\times T_{\omega^{-1}}^d$ associated with
quantum deformations of the $d$-torus and its dual. That
$T_{\omega^{-1}}^d\cong T_\omega^d$ will be a consequence of a set of
Morita equivalences that we shall prove in section 5. In particular, this
observation shows that for {\it any} vertex operator algebra associated
with a positive-definite lattice $L$ of rank $d$, there corresponds a
module over the $d$-dimensional noncommutative torus $T_\omega^d$, with
deformation matrix $\omega$ given as above in terms of the bilinear form of
$L$, which is determined by an $S^1$-twisted projective regular
representation of the group algebra $\complex[L]$. We can summarize these
results in the following

\centerline{ }

{\noindent\baselineskip=12pt
{\bf Proposition 1.} {\em Let $L$ be a positive-definite lattice of rank
$d$ with
bilinear form $g_{ij}$, and let $\alg$ be the vertex operator algebra
associated with $L$. Then the algebra
$\alg^{(\omega)}\subset\alg$ defines a unitary equivalence class of
finitely-generated self-dual $\zed_2$-twisted projective modules
$\widetilde{\cal T}_\omega^d$ of the double noncommutative torus
$T_\omega^{(+)d}\times T_\omega^{(-)d}$ with generators $U_i=V_{e_\pm^i}$
and antisymmetric deformation matrix
\bd
\omega^{ij}=\left\{\new{\begin{array}{rrl}{\rm sgn}(j-i)\,g^{ij}~~&,&~~i\neq
j\\0~~&,&~~i=j\end{array}}\right.
\ed}}

As mentioned in section 1, there is a nice heuristic interpretation of the
noncommutativity of $T_\omega^d$ in the present case. The ordinary torus $T_d$
is formally obtained from $T^d_\omega$ by eliminating the deformation parameter
matrix $\omega\to0$. In the case at hand, $\omega\sim g^{-1}$, so that formally
we let the metric $g\to\infty$ become very large.
As all distances in this formalism are evaluated in terms of a fundamental
length which for simplicity we have set equal to 1, this simply means that at
distance scales larger than this unit of length (which is usually identified
with the Planck length), we recover ordinary, classical spacetime $T_d$. Thus
the classical limit of this quantum deformation of the Lie algebra
$u(1)_+^d\oplus u(1)_-^d$ coincides with the decompactification limit in which
the dual coordinates $x^*$, representing the windings of the string around the
spacetime, delocalize and become unobservable. At very short distances
($g\to0$), spacetime becomes a noncommutative manifold with all of the
exotic duality symmetries implied by string theory. This representation of
the noncommutative torus thus realizes old ideas in string theory about
the nature of spacetime below the fundamental length scale $l_s$
determined by the finite size of the string. In this case $l_s$ is
determined by the lattice spacing of $L$. Therefore, within the framework
of toroidally compactified string theory, spacetime at very short length
scales is a noncommutative torus.

\newsection{Spectral Geometry of Toroidal Compactifications}\label{se:sgtc}

{}From the point of view of the construction of a `space', the pair
$(\alg,{\cal H})$, i.e. a $*$-algebra $\alg$ of operators acting on a
Hilbert space ${\cal H}$, determines only the topology and differentiable
structure of the `manifold'. To put more structure on the space, such as an
orientation and a metric, we need to construct a larger set of data. This
is achieved by using, for even-dimensional spaces, an even real spectral
triple $(\alg,{\cal H},D,J,\Gamma)$ \cite{connesauto}, where $D$ is a
generalized Dirac operator acting on $\cal H$ which determines a Riemannian
structure, $\Gamma$ defines a Hochschild cycle for the geometry which
essentially determines an orientation or Hodge duality operator, and $J$
determines a real structure for the geometry which is used to define
notions such as Poincar\'e duality. In this section we will construct a set
of spectral data to describe a particular noncommutative geometry
appropriate to the spacetime implied by string theory.

To define $(\alg,{\cal H},D,J,\Gamma)$, we need to introduce spinors. For
this, we fix a spin structure on $T_d\times T_d^*$ and
augment the $L^2$ space in \eqn{hilbertdef} to the space $L^2(T_d\times
T_d^*,S)$ of square integrable spinors, where $S\to T_d\times T_d^*$ is the
spin bundle which carries an irreducible left action of the Clifford bundle
$Cl(T_d\times T_d^*)$. We shall take as ${\cal H}$ the corresponding augmented
Hilbert space of \eqn{hilbertdef}, so that, in the notation of the
previous section (see \eqn{H0def}), ${\cal H}_0=L^2(T_d\times T_d^*,S)$. A
dense subspace of $L^2(T_d\times T_d^*,S)$ is provided by the smooth
spinor module ${\cal S}=C^\infty(T_d\times T_d^*,S)$ which is an irreducible
Clifford module of rank $2^d$ over $C^\infty(T_d\times T_d^*)$. There is a
$\zed_2$-grading ${\cal S}={\cal S}^+\oplus{\cal S}^-$ arising from the
chirality grading of $Cl(T_d\times T_d^*)$ by the action of a grading operator
$\Gamma$ with $\Gamma^2=\id$. As a consequence, the augmented Hilbert space
splits as
\beq
{\cal H}={\cal H}^+\oplus{\cal H}^-
\label{hilbertspin}\eeq
into the $\pm1$ (chiral-antichiral) eigenspaces of $\Gamma$.

The spinor module $\cal S$ carries a representation of the double
toroidal Clifford algebra
\beq
\left\{\gamma^\pm_i,\gamma_j^\pm\right\}=\pm\,2g_{ij}~~~~,
{}~~~~\left\{\gamma^\pm_i,
\gamma_j^\mp\right\}=0~~~~,~~~~\left(\gamma^\pm_i\right)^\dagger=
\gamma^\pm_i
\label{gammadefs}\eeq
The $\zed_2$-grading is determined by the chirality matrices
\beq
\Gamma_c^\pm=\mbox{$\frac1{d!\,\sqrt{\det g}}$}\,\epsilon^{i_1i_2\cdots
i_d}\,\gamma_{i_1}^\pm\gamma_{i_2}^\pm\cdots\gamma_{i_d}^\pm
\label{Gammadefs}\eeq
where $\epsilon^{i_1\cdots i_d}$ is the antisymmetric tensor with the
convention $\epsilon^{12\cdots d}=+1$. The matrices $\Gamma_c^\pm$ have the
properties
\beq
\Gamma_c^\pm\gamma_i^\pm=-(-1)^d\,\gamma_i^\pm\Gamma_c^\pm~~~~,~~~~
\Gamma_c^\pm\gamma_i^\mp=(-1)^d\,\gamma_i^\mp\Gamma_c^\pm
{}~~~~,~~~~\left(\Gamma_c^\pm\right)^2=(-1)^{d(d-1)/2}\,\id
\label{Gammaids}\eeq
so that the Klein operator
\beq
\Gamma\equiv(-1)^d\,\Gamma_c^+\Gamma_c^-
\label{kleinop}\eeq
satisfies the desired properties for a grading operator. The
representation of the vertex
operator algebra $\alg$ on $\cal H$ acts diagonally with respect to the
chiral decomposition \eqn{hilbertspin}, i.e.
\beq
\left[\Gamma,V\right]=0~~~~~~\forall V\in\alg
\label{diagaction}\eeq

It was shown in \cite{lslong} that the two natural Dirac operators $D$
associated with a lattice vertex operator algebra are the fields
\beq
\Dirac^\pm=\gamma_i^\pm\otimes z_\pm\alpha_\pm^i(z_\pm)
\label{Diracdefs}\eeq
They are self-adjoint operators on $\cal H$ with compact resolvent on an
appropriate dense domain (after regularization). They each act off-diagonally
on \eqn{hilbertspin} taking ${\cal H}^\pm\to{\cal H}^\mp$,
\beq
\left\{\Gamma,\Dirac^\pm\right\}=0
\label{diracaction}\eeq
We shall use the chirally symmetric and antisymmetric self-adjoint combinations
\beq
\Dirac=\Dirac^++\Dirac^-~~~~~~,~~~~~~\overline{\Dirac}=\Dirac^+-\Dirac^-
\label{chiralcombs}\eeq

The final quantity we need is the operator $J$, which is needed to define a
real structure. Since the representation of $\alg$ on $\bf h$ is faithful
(after regularization), we naturally have an injective map $\alg\to{\bf h}$
defined
by
\beq
V\mapsto|\psi_V\rangle\equiv\lim_{z_\pm\to0}V(z_+,z_-)|0;0\rangle\otimes
|0\rangle_+\otimes|0\rangle_-
\label{injmap}\eeq
so that $V(\psi_V;z_+,z_-)\equiv V(z_+,z_-)$. This means that the (unique)
vacuum state $|0;0\rangle\otimes|0\rangle_+\otimes|0\rangle_-$ is a cyclic
separating vector for $\bf h$ (by the operator-state correspondence), and we
can therefore define an antilinear self-adjoint unitary isometry $J_c:{\cal
H}\to{\cal H}$ by
\beq
J_c|\psi_V\rangle=|\psi_{V^\dagger}\rangle~~~~~~,~~~~~~J_c\gamma_i^\pm=\pm
\gamma_i^\pm J_c
\label{J0def}\eeq
Then $J$ is defined with respect to the decomposition \eqn{hilbertspin} by
acting off-diagonally as $J_c$ on ${\cal H}^+\to{\cal H}^-$ and as $(-1)^dJ_c$
on ${\cal H}^-\to{\cal H}^+$. Note that on spinor fields $\psi\in L^2(T_d\times
T_d^*,S)\subset{\cal H}$ the action of the operator $J$ is given by
\beq
J\psi={\cal C}\,\overline{\psi}
\label{JH0}\eeq
where $\cal C$ is the charge conjugation matrix acting on the spinor indices.

The antilinear unitary involution $J$ satisfies the commutation relations
\beq
J\Dirac=\Dirac J~~~~,~~~~
J\overline{\Dirac}=\overline{\Dirac}J~~~~,~~~~J\Gamma=
(-1)^d\,\Gamma J~~~~,~~~~J^2=\epsilon(d)\,\id
\label{Jcommops}\eeq
and
\beq
\left[V,JW^\dagger
J^{-1}\right]=0~~~~~~,~~~~~~\left[[\Dirac,V],JW^\dagger
J^{-1}\right]=\left[[\overline{\Dirac},V],JW^\dagger J^{-1}\right]=0
\label{Jcommalg}\eeq
for all $V,W\in\alg$. The mod 4 periodic function $\epsilon(d)$
is given by \cite{connesreal}
\beq
\epsilon(d)=(1,-1,-1,1)
\label{modperfns}\eeq
The dimension-dependent $\pm$ signs in the definition of $J$ arise from
the structure of real Clifford algebra representations. The first condition in
\eqn{Jcommalg} implies that for all $V\in\alg$, $JV^\dagger J^{-1}$ lies in the
commutant $\alg'$ of $\alg$ on $\cal H$, while the second condition is a
generalization of the statement that the Dirac operators are first-order
differential operators. The algebra $\alg'$ defines an anti-representation of
$\alg$, and $J$ can be thought of as a charge conjugation operator.

The spectral data $(\alg,{\cal H},D,J,\Gamma)$, with the relations above and
$D$ taken to be either of the two Dirac operators \eqn{chiralcombs}, determines
an even real spectral triple for the geometry of the space we shall work with.
As shown in \cite{lslong}, the existence of two ``natural" Dirac operators
(metrics) for the noncommutative geometry is not an ambiguous property, because
the two corresponding spectral triples are in fact unitarily equivalent, i.e.
there exist many unitary isomorphisms $U:{\cal H}\to{\cal H}$ which define
automorphisms of the vertex operator algebra ($U\alg U^{-1}=\alg$) and for
which
\beq
U\Dirac=\overline{\Dirac}\,U
\label{unitaryequiv}\eeq
This means that the two spaces are naturally isomorphic at the level of their
spectral triples,
\beq
\Bigl(\alg\,,\,{\cal
H}\,,\,\Dirac\,,\,J\,,\,\Gamma\Bigr)~\cong~\left(\alg\,,\,{\cal
H}\,,\,\overline{\Dirac}\,,\,UJU^{-1}\,,\,U\Gamma U^{-1}\right)
\label{spectripiso}\eeq
and so the two Dirac operators determine the same geometry. This feature, along
with the even-dimensionality of the spectral triple, will have important
consequences in what follows. An isomorphism of the form \eqn{spectripiso} is
called a duality symmetry of the noncommutative string spacetime \cite{lslong}.

We now consider the algebra
$\alg^{(\omega)}$ defined in the last section, and restricted to the tachyon
Hilbert space in terms of the orthogonal projection ${\cal P}_0:{\bf
h}\to{\bf h}_0$ onto the subspace \eqn{H0def}. The
corresponding restrictions of the Dirac operators
\eqn{Diracdefs},\eqn{chiralcombs} are
\beq\new{\begin{array}{l}
\partial\slash~\equiv~{\cal P}_0\,\Dirac\,{\cal
P}_0~=~\mbox{$\frac12$}\left[g^{ij}\left(\gamma_i^++\gamma_i^-\right)\otimes
p_j+\left(\gamma_i^+-\gamma_i^-\right)\otimes
w^i\right]\\\overline{\partial\slash}~\equiv~{\cal
P}_0\,\overline{\Dirac}\,{\cal
P}_0~=~\mbox{$\frac12$}\left[g^{ij}\left(\gamma_i^+-\gamma_i^-\right)\otimes
p_j+\left(\gamma_i^++\gamma_i^-\right)\otimes w^i\right]\end{array}}
\label{Diractachyon}\eeq
Note that the operators $J$ and $\Gamma$ preserve both $\alg^{(\omega)}$ and
${\cal
H}_0$. We can therefore define subspaces $(\alg^{(\omega)},{\cal
H}_0,\partial\slash,J,\Gamma)$ and $(\alg^{(\omega)},{\cal
H}_0,\overline{\partial\slash},J,\Gamma)$ of the geometries represented by the
(isomorphic) spectral triples in \eqn{spectripiso}.

\newsection{Duality Transformations and Morita Equivalence}\label{se:dtme}

We shall now describe some basic features of the twisted modules constructed in
section 3. An important property of $\widetilde{\cal T}_\omega^d$ is that there
is no distinction between the torus $T_d$ and its dual $T_d^*$ within the
tachyon algebra. These two commutative subspaces of the tachyon spacetime are
associated with the subspaces
\beq\new{\begin{array}{l}
{\bf
h}_0^{(-)}~\equiv~\bigotimes_{i=1}^d\ker\left(\alpha_+^i\otimes\id+\id\otimes
\alpha_-^i\right)~\cong~L^2\left(T_d\,,\,\mbox{$\prod_{i=1}^d\frac{dx^i}
{2\pi}$}\right)\\{\bf
h}_0^{(+)}~\equiv~\bigotimes_{i=1}^d\ker\left(\alpha_+^i\otimes\id-\id\otimes
\alpha_-^i\right)~\cong~L^2\left(T_d^*\,,\,\mbox{$\prod_{i=1}^d\frac{dx^*_i}
{2\pi}$}\right)\end{array}}
\label{lowensubs}\eeq
of \eqn{H0def}. The subspace ${\bf h}_0^{(\pm)}$ is the projection of ${\bf
h}_0$ onto those states $|q^+;q^-\rangle$ with $q^+=\pm q^-$ (equivalently
$v=0$ and $q=0$, respectively), and it contains only those highest-weight
modules which occur in complex conjugate pairs of left-right representations of
the current algebra $\widehat{u(1)_+^d}\oplus\widehat{u(1)_-^d}$. From
\eqn{subalggens} we see that if ${\cal P}_0^{(\pm)}:{\bf h}_0\to{\bf
h}_0^{(\pm)}$ are the respective orthogonal projections, then the
corresponding
vertex operator subalgebras are
\beq\new{\begin{array}{l}
{\cal A}^{(-)}~\equiv~{\cal P}_0^{(-)}\,\alg^{(\omega)}\,{\cal
P}_0^{(-)}~\cong~
C^\infty(T_d)\\{\cal A}^{(+)}~\equiv~{\cal P}_0^{(+)}\,\alg^{(\omega)}\,{\cal
P}_0^{(+)}~\cong~ C^\infty(T_d^*)\end{array}}
\label{commsubalgs}\eeq
and represent ordinary (commutative) spacetimes. If we choose a spin structure
on $T_d\times T_d^*$ such that the spinors are periodic along the elements of a
homology basis, then the associated spin bundle is trivial
and the corresponding Dirac operators \eqn{Diractachyon} are
\beq
\partial\slash=-ig^{ij}\gamma_i\otimes\mbox{$\frac\partial{\partial
x^j}$}~~~~~~,~~~~~~\overline{\partial\slash}=-i\gamma_i\otimes
\mbox{$\frac\partial{\partial x_i^*}$}
\label{Diracduals}\eeq
where we have used the coordinate space representations
$p_i=-i\frac\partial{\partial x^i}$ on
$L^2(T_d,\prod_{i=1}^d\frac{dx^i}{2\pi})$ and $w^i=-i\frac\partial{\partial
x_i^*}$ on $L^2(T_d^*,\prod_{i=1}^d\frac{dx^*_i}{2\pi})$ given by the adjoint
actions \eqn{zeromodequant}. We have also introduced the new Dirac matrices
$\gamma_i=\frac12(\gamma^+_i+\gamma^-_i)$ and
$\gamma^i_*=\frac12g^{ij}(\gamma^+_j-\gamma^-_j)$, and we used the fact that
$\mbox{$\frac\partial{\partial x^j}$}$ is zero on $L^2(T^*_d)$ (and an
analogous statement involving $\mbox{$\frac\partial{\partial x^*_j}$}$ and
$T_d$). The Dirac operators \eqn{Diracduals} represent the canonical geometries
of the (noncommutative) torus and its dual.

Thus, as subspaces of the tachyon algebra, tori are identified with their
duals, since, as mentioned in the previous section, there exists a unitary
isomorphism, that interchanges $p\leftrightarrow w$ and $g\leftrightarrow
g^{-1}$, of the Hilbert space $\cal H$ which exchanges the two subspaces in
\eqn{lowensubs} and \eqn{commsubalgs} and also the two Dirac operators
\eqn{Diracduals}. Since this isomorphism does not commute with the projection
operators ${\cal P}_0^{(\pm)}$, distinct classical spacetimes are identified.
The equivalence of them from the point of view of noncommutative geometry is
the celebrated $T$-duality symmetry of quantum string theory. Different choices
of spin structure on $T_d$ induce twistings of the spin bundle, and, along with
some modifications of the definitions of the subspaces \eqn{lowensubs} and
\eqn{commsubalgs}, they induce projections onto different dual tori with
appropriate modifications of \eqn{Diracduals}. But, according to
\eqn{spectripiso}, these spectral geometries are all isomorphic \cite{lslong}.

We now turn to a more precise description of the isomorphism classes in
$\widetilde{\cal T}_\omega^d$. We note first of all that the string geometry is
{\it not} contained entirely within the tachyon sector of the vertex operator
algebra $\alg$. This is because the explicit inner automorphisms of $\alg$
which implement the duality symmetries are constructed from higher-order
perturbations of the tachyon sector (by, for example, graviton operators). The
basic tachyon vertex operators $V_i$ together with the Heisenberg fields
$\alpha_\pm^i(z_\pm)=-iV(\id\otimes\alpha_{-1}^{(\pm)i};1,z_\pm)$
generate an affine Lie group $\widehat{\rm Inn}{ }^{(0)}(\alg)$ of inner
automorphisms of the vertex operator algebra which is in general an enhancement
of the generic affine $U(1)_+^d\times U(1)_-^d$ gauge symmetry \cite{lscastro}.
This property is crucial for the occurrence of string duality as a gauge
symmetry of the noncommutative geometry, and the isomorphisms described above
only occur when the relevant structures are embedded into the full spectral
triple $(\alg,{\cal H},D,J,\Gamma)$.

It is well known that various noncommutative tori with different deformation
parameters are equivalent to each other (For completeness, a brief summary of
the general definition and relevance of Morita equivalence for $C^*$-algebras
is presented in appendix A). For instance, when $d=2$ it can be
shown that the abelian group $\zed\oplus\omega\zed$ is an isomorphism invariant
of $\alg^{(\omega)}$. This means, for example, that the tori $T_{\omega}^2\cong
T_{-\omega}^2\cong T_{\omega+1}^2$ are unitarily equivalent. Moreover, it can
be shown in this case that the algebras $\alg^{(\omega)}$ and
$\alg^{(\omega')}$ are Morita equivalent if and only if they are related by a
discrete M\"{o}bius
transformation \cite{Rieffel}
\beq
\omega'= \frac{a\omega+b}{c\omega+d}~~~~~~,~~~~~~\pmatrix{a&b\cr
c&d\cr}\in
GL(2,\zed)
\label{Moritatori}\eeq
where $GL(2,\zed)$ is the group of $2\times 2$ integer-valued matrices with
determinant $\pm 1$. This then also identifies the tori
$T_\omega^2\cong T_{-1/\omega}^2$. Another copy of the discrete group
$SL(2,\zed)$ appears naturally by requiring that the transformation
\beq
U_1\mapsto U_1^aU_2^b~~~~~~,~~~~~~U_2\mapsto U_1^cU_2^d
\label{sl2zsymm}\eeq
be an automorphism of $\alg^{(\omega)}$. Indeed, the transformation
\eqn{sl2zsymm} preserves the product \eqn{cocycletorus}, and so extends to an
(outer) automorphism of $\alg^{(\omega)}$, if and only if $a d - b c =
\det
{\scriptstyle
 \addtolength{\arraycolsep}{-.5\arraycolsep}
 \renewcommand{\arraystretch}{0.5}
 \left( \begin{array}{cc}
 \scriptstyle a & \scriptstyle b \\
 \scriptstyle c  & \scriptstyle d \end{array} \scriptstyle\right)}
= 1$.

The duality symmetries of the string spacetime described above can be used to
identify certain isomorphisms among the modules $\widetilde{\cal T}_\omega^d$.
These incorporate the usual isomorphisms between noncommutative tori described
above as gauge symmetries, and they also naturally contain, for any $d$, the
discrete geometrical automorphism group $SL(d,\zed)$ of $T_d$. When
the algebra $\alg^{(\omega)}$ of the noncommutative torus $T_\omega^d$ is
embedded as described in section 3 into the full vertex operator algebra
$\alg$,
these isomorphisms are induced by the discrete inner automorphisms of $\alg$
which generate the isometry group of the Narain lattice,
\beq
{\rm Aut}(\Lambda)=O(d,d;\zed)\supset SL(d,\zed)
\label{AutLambda}\eeq
They act on the metric tensor $g_{ij}$ by matrix-valued M\"obius
transformations and induce an $O(d,d;\zed)$ symmetry on the Hilbert space
acting unitarily on the Dirac operators in the sense of \eqn{unitaryequiv}.
Therefore, with the embedding $\alg^{(\omega)}\hookrightarrow\alg$, a much
larger class of tori are identified by the action of the full duality
group\footnote{As emphasized in \cite{Morita} this $O(d,d;\zed)$ transformation
is only defined on the dense subspace of the vector space of antisymmetric
real-valued tensors $\omega$ where $C\omega+D$ is invertible. We shall
implicitly assume this here.},
\beq
T^d_\omega\cong T^d_{\omega^*}~~~~~~{\rm
with}~~\omega^*=(A\omega+B)(C\omega+D)^{-1}~~~,~~~\pmatrix{A&B\cr C&D\cr}\in
O(d,d;\zed)
\label{nctorusisos}\eeq
where $A,B,C,D$ are $d\times d$ integer-valued matrices satisfying the
relations
\beq
A^\top C+C^\top A=0=B^\top D+D^\top B~~~~~~,~~~~~~A^\top D+C^\top B=\id
\label{ABCDconds}\eeq
To see this, we embed the two algebras $\alg^{(\omega)}$ and
$\alg^{(\omega^*)}$ into $\alg$. Then the unitary equivalence
\eqn{spectripiso} implies that, in $\alg$, there is a finitely-generated
projective right $\alg^{(\omega)}$-module ${\cal E}^{(\omega)}$
with $\alg^{(\omega^*)}\cong{\rm End}_{\alg^{(\omega)}}
{\cal E}^{(\omega)}$ where the $*$-isomorphism is implemented by
the unitary operator $U$ in \eqn{unitaryequiv}.
Thus the  projections back onto these tachyonic algebras establishes the
Morita equivalence \eqn{nctorusisos} between the twisted
modules.\footnote{Strictly speaking, in order to correctly identify
\eqn{nctorusisos} as the orbits under the action of the target space duality
group of the string theory, one must regard $\omega$ as the natural
antisymmetric bilinear form induced on the lattice $\Gamma$ by the metric $g$,
as defined above. This means that the string background in effect has an
induced antisymmetric tensor $B$ which parametrizes the deformation of the
torus, as in \cite{cds,dh} (and thus transforms in the way stated -- see
\cite{lslong} for the details). The action of the duality group in
\eqn{nctorusisos} is then an explicit realization of the duality transformation
law proposed in \cite{cds} and proven in \cite{Morita,schwarz}.}
In other words, the tori in \eqn{nctorusisos}, being Morita
equivalent, are indistinguishable when embedded in $\alg$.

To summarize, string duality implies the

\centerline{ }

{\noindent\baselineskip=12pt
{\bf Proposition 2.} {\em There is a natural Morita equivalence
\bd
T^d_\omega\cong T^d_{\omega^*}~~~~~~;~~~~~~
\omega^*\equiv(A\omega+B)(C\omega+D)^{-1}~~~,~~~
\pmatrix{A&B\cr C&D\cr}\in O(d,d;\zed)
\ed
of spectral geometries.}}

\centerline{ }

\noindent
A similar result for multi-dimensional noncommutative tori has been established
recently in \cite{Morita} using more explicit formal constructions of
projective modules. Here we see the power of duality in establishing this
strong result. It is intriguing that the symmetry group \eqn{AutLambda}
contains the $SL(2,\zed)$ $S$-duality symmetry of type-IIB superstring theory.
In \cite{schwarz} it is shown that Matrix Theory compactifications on Morita
equivalent noncommutative tori are physically equivalent, in that the
associated quantum theories are dual. Here we find a similar manifestation of
this property, directly in the language of string geometry. The construction in
the present case exposes the natural relationship between target space duality
and Morita equivalence of noncommutative geometries. It also demonstrates
explicitly in what sense compactifications on Morita equivalent tori are
physically equivalent, as conjectured in \cite{cds}. For instance, the
equivalence $\omega\leftrightarrow\omega^{-1}$ represents the unobservability
of small distances in the physical spacetime, while the equivalence
$\omega\leftrightarrow\omega+1$ (for $d=2$) represents the invariance of the
spacetime under a change of complex structure.

As we discussed above, string duality can be represented as a gauge symmetry on
the full vertex operator algebra $\alg$ and, when projected onto the tachyon
sector, one obtains intriguing equivalences between the
twisted realizations of the noncommutative torus. In this representation of
$T_\omega^d$, the automorphisms of the algebra $\alg^{(\omega)}$ are determined
in large part by gauge transformations, i.e. the elements of $\widehat{\rm
Inn}{ }^{(0)}(\alg)$,\footnote{This Lie group is described in detail in
\cite{lscastro}.} consistent with the corresponding results for the ordinary
noncommutative torus. In particular, the usual diffeomorphism symmetries are
given by inner automorphisms, i.e. gravity becomes a gauge theory on this
space. The outer automorphisms of $\alg$ are given by the full duality group of
toroidally compactified string theory which is the semi-direct product
$O(d,d;\zed)~\semiprod O(2,\real)$, where $O(2,\real)$ is a worldsheet symmetry
group that acts on the algebra $\alg$ by rotating the two chiral sectors on
$\complex\cup\{\infty\}$ among each other (this part of the duality group does
not act on the metric tensor of $T_d$). After applying the appropriate
projections, we obtain canonical actions of these automorphisms on the tachyon
sector. Thus, again using string duality, we arrive at the

\centerline{ }

{\noindent\baselineskip=12pt
{\bf Proposition 3.} {\em There is a natural subgroup of automorphisms of
the
twisted module $\widetilde{\cal T}^d_\omega$ given by the duality group
\bd
{\rm Aut}^{(0)}(\alg^{(\omega)})={\cal P}_0\left(\widehat{\rm Inn}{
}^{(0)}(\alg)~\semiprod{\rm Out}(\alg)\right){\cal P}_0
\ed
where $\widehat{\rm Inn}{ }^{(0)}(\alg)$ is the affine Lie group generated by
the tachyon vertex operators $V_i$ and the Heisenberg fields $\alpha_\pm^i$,
and ${\rm Out}(\alg)\cong O(d,d;\zed)~\semiprod O(2,\real)$.}}

\newsection{Gauge Theory on $\widetilde{\cal T}^d_\omega$}

In noncommutative geometry a finitely-generated projective module of an algebra
replaces the classical notion of a vector bundle over a manifold
\cite{Book,giannibook}. The geometry of gauge theories can therefore also be
cast into the natural algebraic framework of noncommutative geometry. In this
section we will construct the gauge theories on $\widetilde{\cal
T}^d_\omega$ which in the next section will enable us to construct an action
functional that is naturally invariant under the automorphism group given by
Proposition 3, and in particular under duality and diffeomorphism
symmetries.
The possibility of such a characterization comes from the fairly complete
mathematical theory of projective modules and of connections on these modules
for the noncommutative torus $T_\omega^d$ \cite{ncmodules}.

On $\alg^{(\omega)}$ there is a natural set of linear derivations $\Delta_i$,
$i=1,\dots,d$, defined by
\beq
\Delta_i(U_j)=\delta_{ij}\,U_j
\label{diffderivs}\eeq
These derivations come from the derivative operators
$p_i=-i\frac\partial{\partial x^i}$ acting on the algebra $C^\infty(T_d)$
of Fourier series (see section 3). They generate a $u(1)^d$ Lie algebra.
There is a natural representation of the operators $\Delta_i$ in the
twisted module. The tachyon vertex operators $V_{q^+q^-}(z_+,z_-)$ have
charge $g^{ij}q_j^\pm$ under the action of the $u(1)_\pm^d$ current algebra
generated by the Heisenberg fields \eqn{heisfieldsdef}. As such
$\alpha_\pm^i$ generate an affine Lie algebra $\widehat{\Lambda^c}$ of
automorphisms of $\alg^{(\omega)}$ (Proposition 3). We can define a set of
linear derivations $\nabla_\pm^{(0)i}$, $i=1,\dots,d$, acting on
$\alg^{(\omega)}$ by the infinitesimal adjoint action of the Heisenberg
fields \cite{lslong,lscastro} which is given by
\beq
\nabla_\pm^{(0)i}\,V_{e^\pm_j}\equiv\left[\alpha_\pm^i,V_{e^\pm_j}\right]
=\delta^i_j\,V_{e^\pm_j}
\label{derivdefs}\eeq

This leads us to the notion of a connection on the algebra
$\alg^{(\omega)}$.
We shall start by considering this algebra as a finitely-generated
projective left module over itself. A (chiral) left connection on
$\alg^{(\omega)}$ is then defined as a set of $\complex$-linear operators
$\nabla_\pm^i$, $i=1,\dots,d$, acting on $\alg^{(\omega)}$ and satisfying the
left
Leibniz rule
\beq
\nabla_\pm^i(VW)=V\,\nabla_\pm^iW+\left[\alpha_\pm^i,V
\right]W~~~~~~,~~~~~~\forall V,W\in\alg^{(\omega)}
\label{Leibniz}\eeq
If $\nabla_\pm^i$ and $\nabla_\pm^{\prime i}$ are two connections on
$\alg^{(\omega)}$,
then their difference $\nabla_\pm^i-\nabla_\pm^{\prime i}$ commutes with the
left action of $\alg^{(\omega)}$ on itself, i.e.
$(\nabla_\pm^i-\nabla_\pm^{\prime
i})(VW)=V(\nabla_\pm^i-\nabla_\pm^{\prime i})W$. Thus
$\nabla_\pm^i-\nabla_\pm^{\prime i}$ is an element of the endomorphism
algebra ${\rm End}_{\alg^{(\omega)}}\alg^{(\omega)}$ which we identify
with the commutant ${\rm End}_{\alg^{(\omega)}}\alg^{(\omega)}
\cong{\alg^{(\omega)}}' = J\alg^{(\omega)}J^{-1}$ of the algebra
$\alg^{(\omega)}$, where $J$ is the real structure introduced in section 4.
The derivations defined in \eqn{derivdefs} also satisfy the Leibnitz rule
\eqn{Leibniz} and thus define fixed fiducial elements in the space of
connections on $\alg^{(\omega)}$. It follows that
$\Omega_\pm^i=\nabla_\pm^i-\nabla_\pm^{(0)i}$ obeys
\beq
\Omega_\pm^iVW=V\,\Omega_\pm^iW~~~~~~,~~~~~~\forall V,W\in\alg^{(\omega)}
\label{Apmcommprop}\eeq
so that we can write an arbitrary connection in the form
\beq
\nabla_\pm^i=\nabla_\pm^{(0)i}+\Omega_\pm^i~~~~~~{\rm
with}~~\Omega_\pm^i\in{\rm End}_{\alg^{(\omega)}}\alg^{(\omega)}
\label{connpar}\eeq

We can introduce a Hermitian structure on $\alg^{(\omega)}$ via the
$\alg^{(\omega)}$-valued
positive-definite inner product
$\langle\cdot~,~\cdot\rangle_{\alg^{(\omega)}}:\alg^{(\omega)}\times
\alg^{(\omega)}\to\alg^{(\omega)}$ defined by
\beq
\langle V,W\rangle_{\alg^{(\omega)}}=V^\dagger
W~~~~~~,~~~~~~V,W\in\alg^{(\omega)}
\label{Hermstrucdef}\eeq
The compatibility condition with respect to this Hermitian structure for left
connections,
\beq
-\left\langle\nabla_\pm^iV,W\right\rangle_{\alg^{(\omega)}}+\left\langle
V,\nabla_\pm^iW\right\rangle_{\alg^{(\omega)}}=\nabla_\pm^{(0)i}\Bigl(\langle
V,W\rangle_{\alg^{(\omega)}}\Bigr)
\label{compcondn}\eeq
implies that $\Omega_\pm^i=(\Omega_\pm^i)^\dagger$ is self-adjoint.
The minus sign in \eqn{compcondn} arises from the fact that the fiducial
connection defined in \eqn{derivdefs} anticommutes with the $*$-involution,
$(\nabla_\pm^{(0)i} V)^\dagger = -\nabla_\pm^{(0)i} V^ \dagger$.

Since the connection coefficients $\Omega_\pm^i$ in \eqn{connpar} are elements
of the commutant ${\alg^{(\omega)}}' \cong {\rm End}_{\alg^{(\omega)}}
\alg^{(\omega)}$, it is natural to
introduce an $\alg^{(\omega)}$-bimodule structure. We first define a right
$\alg^{(\omega)}$-module which we denote by $\overline{\alg^{(\omega)}}$.
Elements of
$\overline{\alg^{(\omega)}}$ are in bijective correspondence with those of
${\alg^{(\omega)}}$, $\overline{\alg^{(\omega)}} \equiv \{\overline{V} ~|~
V \in \alg^{(\omega)} \}$, and
the right action of $V \in {\alg^{(\omega)}}$ on $\overline{W} \in
\overline{\alg^{(\omega)}}$ is
given by
$\overline{W} \cdot V = \overline{V^\dagger W}$. Associated with the connection
$\nabla^i_\pm$ on $\alg^{(\omega)}$ there is then a right connection
$\overline{\nabla}^i_\pm$ on $\overline{\alg^{(\omega)}}$ defined by
\beq
\overline{\nabla}^i_\pm \overline{V} = - \overline{ \nabla^i_\pm V}
{}~~~~~~,~~~~~~\forall \overline{V} \in \overline{\alg^{(\omega)}}
\label{rightconndef}\eeq
The operator \eqn{rightconndef} is $\complex$-linear and obeys a right Leibniz
rule. Indeed, with $\overline{V} \in \overline{\alg^{(\omega)}}$ and $W \in
{\alg^{(\omega)}}$,
we have
\beq\new{\begin{array}{lll}
\overline{\nabla}^i_\pm (\overline{V} \cdot W) &=&
\overline{\nabla}^i_\pm (\overline{W^\dagger V})\\&=& -
\overline{\nabla^i_\pm ( W ^\dagger V)}\\&=& -
\overline{ W^\dagger \nabla^i_\pm V} -
\overline{\left(\nabla_\pm^{(0)i}W^\dagger\right) V}\\&=& -
\overline{\nabla^i_\pm V} \cdot W +
\overline{\left(\nabla_\pm^{(0)i}W\right) ^\dagger V} \\&=&
\left(\overline{\nabla}^i_\pm \overline{V}\right) \cdot W
+\overline{V} \cdot \left[\alpha^i_\pm,W\right]\end{array}}
\label{rightleibniz}\eeq
which expresses the right Leibniz rule for $\overline{\nabla}^i_\pm$.

We can now combine the two connections $\nabla^i_\pm$ and
$\overline{\nabla}^i_\pm$ to get a symmetric connection
$\widetilde{\nabla}^i_\pm$ on $\overline{\alg^{(\omega)}}\otimes_{
\alg^{(\omega)}}\alg^{(\omega)}$ defined
by
\beq
\widetilde{\nabla}^i_\pm(\overline{V}\otimes
W)=\overline{\nabla}^i_\pm\overline{V}\otimes W+\overline{V}
\otimes\nabla^i_\pm W
{}~~~~~~,~~~~~~\forall \overline{V}\otimes W \in
\overline{\alg^{(\omega)}}\otimes_{\alg^{(\omega)}}\alg^{(\omega)}
\label{totconndef}\eeq
The $\alg^{(\omega)}$-bimodule we are describing is just the
Hilbert space ${\cal H}_0$ which carries both a left representation of
$\alg^{(\omega)}$ and a (right) anti-representation of
$\alg^{(\omega)}$ given by
$V\mapsto JV^\dagger J^{-1} \in {\cal L}({\cal H}_0)$, where
${\cal L}({\cal H}_0)$ is the algebra of bounded linear operators on
${\cal H}_0$. In the present interpretation, the right
module structure of ${\cal H}_0$ comes from the right $\alg^{(\omega)}$-module
$\overline{\alg^{(\omega)}}$ when mapping
$\overline{\alg^{(\omega)}}$ into ${\cal
L}({\cal H}_0)$
by
$\overline{V} \mapsto JV^\dagger J^{-1}$. Thus, when representing
$\overline{\alg^{(\omega)}}\otimes_{\alg^{(\omega)}}\alg^{(\omega)}$
on ${\cal H}_0$, using
\eqn{connpar}
and \eqn{rightconndef} we find that the action of the connection
\eqn{totconndef} restricted to $\alg^{(\omega)}$ can be expressed in terms
of the
fiducial connection \eqn{derivdefs} and the connection coefficients as
\beq
\widetilde{\nabla}_\pm^i(\id\otimes
V)=\id\otimes\left(\nabla_\pm^{(0)i}+\Omega_\pm^i+J(\Omega_\pm^i)^\dagger
J^{-1}\right)V
\label{connreal}\eeq
The extra connection term in \eqn{connreal} achieves the desired left-right
symmetric representation and can be thought of as enforcing $CPT$-invariance.

The operator
\beq
\widetilde{\nabla\slashs}^{\,\pm}\equiv\gamma_i^\pm\otimes
\widetilde{\nabla}^i_\pm
\label{nablaslashdef}\eeq
is then a map on
$\overline{\alg^{(\omega)}}\otimes_{\alg^{(\omega)}}\alg^{(\omega)}\to
\overline{\alg^{(\omega)}}\otimes_{\alg^{(\omega)}}
\Omega^1_{\Dirac^\pm}(\alg^{(\omega)})\otimes_{\alg^{(\omega)}}
\alg^{(\omega)}$, where
\beq
\Omega_{\Dirac^\pm}^1(\alg^{(\omega)})={\rm
span}_\complexs\left\{V\left[\Dirac^\pm,W\right]~\Bigm|~V,W\in
\alg^{(\omega)}\right\}
\label{1formszerodef}\eeq
are linear spaces of one-forms which carry a natural $\alg^{(\omega)}$-bimodule
structure. The action of the operator $\widetilde{\nabla\slashs}^{\,\pm}$
on $\alg^{(\omega)}$ as defined in \eqn{connreal} can be expressed in terms
of the
adjoint actions of the Dirac operators \eqn{Diracdefs} as
$\widetilde{\nabla\slashs}^{\,\pm}|_{\alg^{(\omega)}}={\rm
Ad}_{\Dirac^\pm}+A^\pm+J(A^\pm)^\dagger J^{-1}$, where
$A^\pm\in\Omega^1_{\Dirac^\pm}(\alg^{(\omega)})$ are self-adjoint operators (by
the compatibility condition \eqn{compcondn}) which are called gauge
potentials. In
fact, all of this structure is induced by a covariant Dirac operator which is
associated with the full data $(\alg,{\cal H},J,\Gamma)$ and is defined as
\beq
\Dirac^\pm_\nabla=\Dirac^\pm+A^\pm+J(A^\pm)^\dagger J^{-1}~~~~~~,
{}~~~~~~A^\pm\in\Omega_{\Dirac^\pm}^1(\alg)
\label{Diraccovdefs}\eeq
where $\Omega_{\Dirac^\pm}^1(\alg)$ are the $\alg$-bimodules of one-forms
defined as in \eqn{1formszerodef} but with the algebra
$\alg^{(\omega)}$
replaced by the full vertex operator algebra $\alg$. The Dirac operator
$\Dirac^\pm_\nabla$ is regarded as an internal perturbation of $\Dirac^\pm$ and
it yields a geometry that is unitary equivalent to that determined by
$\Dirac^\pm$ \cite{connesauto}, i.e. the geometries with fixed data
$(\alg,{\cal
H},J,\Gamma)$ form an affine space modelled on $\Omega_{\Dirac^\pm}^1(\alg)$.

The spaces $\Omega^1_{\Dirac^\pm}(\alg)$ are free $\alg$-bimodules with bases
$\{\gamma_i^\pm\}_{i=1}^d$ \cite{lslong}. The gauge potentials in
\eqn{Diraccovdefs} can therefore be decomposed as
\beq
A^\pm=g^{ij}\gamma_i^\pm\otimes A^\pm_j~~~~~~{\rm with}~~A^\pm_i\in\alg
\label{gaugecomps}\eeq
Defining the self-adjoint elements $A_i\equiv{\cal P}_0(A_i^++A_i^-){\cal P}_0$
and $A^i_*={\cal P}_0\,g^{ij}(A_j^+-A_j^-){\cal P}_0$ of $\alg^{(\omega)}$, the
covariant
versions of the Dirac operators \eqn{Diractachyon}, obtained from the
restrictions of \eqn{Diraccovdefs} to $\alg^{(\omega)}$, are then
\beq\new{\begin{array}{lll}
{\partial\slash}_\nabla~\equiv~
{\cal P}_0\left(\Dirac^{+}_\nabla + \Dirac^{-}_\nabla\right) {\cal P}_0
&=&\mbox{$\frac12$}\left[g^{ij}\left(\gamma_i^++
\gamma_i^-\right)\otimes\left(p_j+A_j+g_{jk}\,JA_*^kJ^{-1}\right)\right.\\&
&~~~~~~\left.+
\left(\gamma_i^+-\gamma_i^-\right)\otimes\left(w^i+A_*^i+g^{ij}JA_jJ^{-1}
\right)\right]\\
\overline{\partial\slash}_\nabla~\equiv~
{\cal P}_0\left(\Dirac^{+}_\nabla - \Dirac^{-}_\nabla\right) {\cal P}_0
&=&\mbox{$\frac12$}
\left[g^{ij}\left(\gamma_i^+-\gamma_i^-\right)\otimes\left(p_j+A_j+g_{jk}\,
JA_*^kJ^{-1}\right)\right.\\&
&~~~~~~\left.+\left(\gamma_i^++\gamma_i^-\right)\otimes\left(w^i+A_*^i+g^{ij}
JA_jJ^{-1}\right)\right]\end{array}}
\label{Dcovtachyon}\eeq
where we have used \eqn{J0def}.

The final ingredient we need for a gauge theory on $\widetilde{\cal
T}_\omega^d$ is some definition of an invariant integration in order to define
an action functional. The trace \eqn{nctorustrace} yields a natural normalized
trace ${\rm Tr}:\alg^{(\omega)}\to\complex$ on the twisted module defined by
\beq
{\rm Tr}\,V=\int_{T_d}\int_{T_d^*}\,
\prod_{i=1}^d\frac{dx^i~dx_i^*}{(2\pi)^2}~V(x,x^*)
\label{Trdef}\eeq
This trace is $\widehat{\Lambda^c}$-invariant, ${\rm
Tr}(\nabla_\pm^{(0)i}V)=0~~\forall V\in\alg^{(\omega)}$, and the corresponding
Gelfand-Naimark-Segal representation space $L^2(\alg^{(\omega)},{\rm Tr})$
is, by the
operator-state correspondence, canonically isomorphic to the Hilbert space
${\bf h}_0$. Using the trace \eqn{Trdef} and the $\alg^{(\omega)}$-valued
inner product
\eqn{Hermstrucdef} we obtain a usual complex inner product
$(\cdot~,~\cdot)_{\alg^{(\omega)}}:\alg^{(\omega)}
\times\alg^{(\omega)}\to\complex$ defined by
\beq
(V,W)_{\alg^{(\omega)}}={\rm Tr}\,\langle
V,W\rangle_{\alg^{(\omega)}}=\left\langle\psi_V|\psi_W\right\rangle_{{\bf h}_0}
\label{innerproddef}\eeq
which coincides with the inner product on the Hilbert space ${\bf h}_0$. Being
functions which are constructed from invariant traces, the quantities
\eqn{Trdef} and \eqn{innerproddef} are naturally invariant under unitary
transformations of the algebra $\alg^{(\omega)}$, and, in particular,
under the
action of the inner automorphism group given in Proposition 3. This
property
immediately implies the manifest duality-invariance of any action functional
constructed from them.

\newsection{Duality-symmetric Action Functional}\label{se:dia}

The constructions of the preceding sections yield a completely
duality-symmetric formalism, and we are now ready to define a manifestly
duality-invariant action functional
\beq
I[\psi,\widehat{\psi};\nabla,\nabla_{\!*}]=I_B[\nabla,\nabla_{\!*}]
+I_F[\psi,\widehat{\psi};\nabla,\nabla_{\!*}]
\label{totaction}\eeq
associated with a generic gauge theory on the twisted module. The bosonic part
of the action is defined as
\beq
I_B[\nabla,\nabla_{\!*}]\equiv\mbox{$\frac1{2^d}$}\,{\rm
Tr}_S\left[\Pi\left({\partial\slash}_\nabla^{\,2}
-\overline{\partial\slash}_\nabla^{\,2}\right)\right]^2
\label{bosactiondef}\eeq
where ${\rm Tr}_S$ is the trace \eqn{Trdef} including a trace over the Clifford
module, and $\Pi$ is the projection operator onto the space of
antisymmetric tensors (two-forms). This projection is equivalent to
the quotienting by junk forms of the representation of the
universal forms
on the Hilbert space \cite{Book,giannibook}. The operator
${\partial\slash}_\nabla^{\,2}-\overline{\partial\slash}_\nabla^{\,2}$ is the
lowest order polynomial combination of the two Dirac operators
\eqn{Dcovtachyon} which lies in the endomorphism algebra ${\rm
End}_{\alg^{(\omega)}}\alg^{(\omega)}$.
Thus the action \eqn{bosactiondef} comes from the lowest
order polynomial multiplication operator two-form
which is invariant under the duality
symmetries represented by an interchange of the Dirac operators. The fermionic
action is of the form of a Dirac action. Let $\psi = (\psi^a)_{a=1}^{2^d}$ be a
square-integrable section of the spin bundle over $T_d\times T_d^*$. Then using
a flat metric $\delta_{ab}$ for the spinor indices, we define
\beq
I_F[\psi,\widehat{\psi};\nabla,\nabla_{\!*}]\equiv
{\rm Tr}\,\sum_{a=1}^{2^d}\left\langle
J^{-1}V(\psi^a)J\,,\,\left[{\partial\slash}_\nabla,V(\psi^a)
\right]\right\rangle_{\alg^{(\omega)}}= \left\langle\widehat{\psi}
\Bigl|\,{\partial\slash}_\nabla\,\Bigr|\psi\right\rangle_{{\bf h}_0}
\label{fermactiondef}\eeq
where
\beq
\widehat{\psi}=J^{-1}\,\psi=\epsilon(d)\,{\cal C}\,\overline{\psi}
\label{hatpsidef}\eeq
is the corresponding anti-fermion field. In \eqn{fermactiondef} $V(\psi^a)$
denotes the map from the Hilbert space into $\alg^{(\omega)}$.

The actions \eqn{bosactiondef} and \eqn{fermactiondef} are both gauge-invariant
and depend only on the spectral properties of the Dirac $K$-cycles $({\cal
H}_0,\partial\slash)$ and $({\cal H}_0,\overline{\partial\slash})$. A gauge
transformation is an inner automorphism $\sigma_u:\alg\to\alg$, parametrized
by a unitary element $u$ of $\alg$, i.e. an element of the unitary group ${\cal
U}(\alg)=\{u\in\alg~|~u^\dagger u=uu^\dagger=\id\}$ of $\alg$, and defined by
\beq
\sigma_u(V)=uVu^\dagger~~~~~~,~~~~~~V\in\alg
\label{innautodef}\eeq
It acts on gauge potentials as
\beq
A^\pm~\mapsto~(A^\pm)^u\equiv uA^\pm
u^\dagger+u\left[\Dirac^\pm,u^\dagger\right]
\label{gaugetransf}\eeq
and, using the $\alg$-bimodule structure on $\cal H$, on spinor fields by the
adjoint representation
\beq
\psi~\mapsto~\psi^u\equiv u\,\psi\,u^\dagger=U\psi
\label{spinortransf}\eeq
where
\beq
U=uJuJ^{-1}
\label{Udef}\eeq
The transformation \eqn{spinortransf} preserves the inner product on the
Hilbert space $\cal H$,
$\langle\psi_1^u|\psi_2^u\rangle=\langle\psi_1|\psi_2\rangle$, since both $u$
and $J$ are isometries of $\cal H$. Moreover, one easily finds that, under a
gauge transformation \eqn{gaugetransf},
\beq
{\nabla\slashs}^{\,\pm}~\mapsto~({\nabla\slashs}^{\,\pm})^u\equiv
u\,{\nabla\slashs}^{\,\pm}
\,u^\dagger~~~~~~,~~~~~~\Dirac^\pm_\nabla~\mapsto~\Dirac^\pm_{\nabla^u}
=U\,\Dirac^\pm_\nabla\,U^\dagger
\label{covderivtransf}\eeq
which immediately shows that
\beq
I[\psi^u,\widehat{\psi}^u;\nabla^u,\nabla_{\!*}^u]=I[\psi,\widehat{\psi};
\nabla,\nabla_{\!*}]
\label{actioninv}\eeq

Let us write the action \eqn{totaction} in a form which makes its duality
symmetries explicit. Using the double Clifford algebra \eqn{gammadefs} and the
coordinate space representations of the momentum and winding operators in
\eqn{Dcovtachyon}, after some algebra we find
\beq\new{\begin{array}{l}
{\partial\slash}_\nabla^{\,2}-\overline{\partial\slash}_\nabla^{\,2} = -\frac
i2\,\gamma_+^i\gamma_-^j\otimes\left(\frac{\partial a_j}{\partial
x^i}-\frac{\partial a_i}{\partial x^j}-g_{ik}g_{jl}\frac{\partial
a_*^l}{\partial x_k^*}+g_{ik}g_{jl}\frac{\partial a_*^k}{\partial
x_l^*}+i\Bigl[a_i,a_j\Bigr]-ig_{ik}g_{jl}\left[a_*^k,a_*^l\right]\right.\\
{}~~~~~~~~~~~~~~~~~~~~~\left.+\,g_{jk}\frac{\partial a_*^k}{\partial
x^i}+g_{ik}\frac{\partial a_*^k}{\partial
x^j}+ig_{jk}\left[a_i,a_*^k\right]-g_{jk}\frac{\partial a_i}
{\partial x^*_k}-g_{ik}\frac{\partial a_j}{\partial x^*_k}+ig_{ik}\left[
a_j,a_*^k\right]\right)\end{array}}
\label{Diracdiff}\eeq
where we have defined $a_i=A_i+g_{ij}JA_*^jJ^{-1}$ and
$a_*^i=A_*^i+g^{ij}JA_jJ^{-1}$. The projection operator $\Pi$ acting on
\eqn{Diracdiff} sends the gamma-matrix product $\gamma_+^i\gamma_-^j$
into its antisymmetric component
$\frac12[\gamma_+^i,\gamma_-^j]$, thus eliminating from
\eqn{Diracdiff} the symmetric part. Since
conjugation of (the components of) a gauge potential by the
real structure $J$ produces elements of the commutant ${\alg^{(\omega)}}'$ (see
\eqn{Jcommalg}), we can write the bosonic action \eqn{bosactiondef} in the form
of a symmetrized Yang-Mills type functional
\beq\new{\begin{array}{lll}
I_B[\nabla,\nabla_{\!*}]&=&\int_{T_d}\int_{T_d^*}\,
\prod_{i=1}^d\frac{dx^i~dx^*_i}{(2\pi)^2}~g^{ik}g^{jl}\left(F_{ij}^{\nabla,
\nabla_{\!*}}F_{kl}^{\nabla,\nabla_{\!*}}+J\,\widetilde{F}_{ij}^{\nabla,
\nabla_{\!*}}\widetilde{F}_{kl}^{\nabla,\nabla_{\!*}}\,J^{-1}\right.\\&
&\left.~~~~~~~~~~~~~~~~~~~~~~~~~~~~~~~~~~~+\,2\,F_{ij}^{\nabla,\nabla_{\!*}}\,J
\,\widetilde{F}_{kl}^{\nabla,\nabla_{\!*}}\,J^{-1}\right)\end{array}}
\label{bosactionexpl}\eeq
where
\bea
F_{ij}^{\nabla,\nabla_{\!*}}&=&\frac{\partial A_j}{\partial x^i}-\frac{\partial
A_i}{\partial x^j}+i\Bigl[A_i,A_j\Bigl]-g_{ik}g_{jl}\left(\frac{\partial
A_*^l}{\partial x^*_k}-\frac{\partial A_*^k}{\partial
x^*_l}+i\left[A_*^k,A_*^l\right]\right)\label{YMcurv}\\
\widetilde{F}_{ij}^{\nabla,\nabla_{\!*}}&=&g_{jk}\frac{\partial A_*^k}{\partial
x^i}-g_{ik}\frac{\partial A_*^k}{\partial
x^j}+ig_{ik}g_{jl}\left[A_*^k,A_*^l\right]\nonumber\\&
&~~~~~~~~~~~~~~~~~~~~~~~~~-\left(g_{ik}\frac{\partial A_j}{\partial
x^*_k}-g_{jk}\frac{\partial A_i}{\partial x^*_k}+i\Bigl[A_i,A_j\Bigr]\right)
\label{YMcurvdual}\eea
Note that the field strength \eqn{YMcurvdual} is obtained from \eqn{YMcurv} by
interchanging the gauge potentials $A_i\leftrightarrow g_{ij}A_*^j$, but {\it
not} the local coordinates $(x,x^*)$. According to the description of
section 3 (see Proposition 1),
the commutators in \eqn{YMcurv} and
\eqn{YMcurvdual} can be defined using the Moyal bracket
\beq
[A,B]\equiv\{A,B\}_{\omega}=A\star_\omega B-B\star_\omega A
\label{moyalbracket}\eeq
where
\beq
(A\star_\omega
B)(x,x^*)=\exp\left[i\pi\omega^{ij}\left(\mbox{$\frac\partial{\partial
x^i}\frac\partial{\partial x'^j}-g_{ik}g_{jl}\frac\partial{\partial
x^*_k}\frac\partial{\partial
x'^*_l}$}\right)\right]A(x,x^*)B(x',x'^*)\biggm|_{(x',x'^*)=(x,x^*)}
\label{starproddouble}\eeq
is the deformed product on $C^\infty(T_d\times T_d^*)$ (see also ref.
\cite{Fairlie}).

The fermionic action can be written in a more transparent form as follows. We
fix a spin structure such that any spinor field on $T_d\times T_d^*$ can be
decomposed into a periodic spinor $\chi$ and an antiperiodic spinor $\chi_*$,
\beq
\psi=\chi\oplus\chi_*
\label{periodicdecomp}\eeq
with respect to a homology basis. They are defined by the conditions
\beq
\left(\gamma_i^+-\gamma_i^-\right)\chi=0~~~~~~,~~~~~~
\left(\gamma_i^++\gamma_i^-\right)\chi_*=0
\label{chidefs}\eeq
for all $i=1,\dots,d$. It is important to note that this periodic-antiperiodic
decomposition is not the same as the chiral-antichiral one in
\eqn{hilbertspin}, although its behaviour under the action of the charge
conjugation operator $J$ is very similar. From \eqn{J0def} it follows that the
corresponding anti-spinors $\widehat{\chi}=J^{-1}\chi$ and
$\widehat{\chi}_*=J^{-1}\chi_*$ obey, respectively, antiperiodic and
periodic conditions. Furthermore, as there are $2^d$ possible choices
of spin structure on the $d$-torus $T_d$, there are many other analogous
decompositions that one can make. However, these choices are all related by
``partial" $T$-duality transformations of the noncommutative geometry
\cite{lslong} and hence the fermionic action is independent of the choice of
particular spin structure. Here we choose the one which makes its duality
symmetries most explicit. Defining, with the usual conventions, the
gamma-matrices $\gamma_i=\frac12(\gamma_i^++\gamma_i^-)$ and
$\gamma_*^i=\frac12g^{ij}(\gamma_j^+-\gamma_j^-)$, after some algebra we find
that the fermionic action \eqn{fermactiondef} can be written in terms of the
decomposition \eqn{periodicdecomp} as
\beq\new{\begin{array}{l}
I_F[\chi,\chi_*,\widehat{\chi},\widehat{\chi}_*;\nabla,\nabla_{\!*}]\\~~~~~~=~
\int_{T_d}\int_{T_d^*}\,\prod_{i=1}^d\frac{dx^i~dx^*_i}{(2\pi)^2}~
\left[-i\widehat{\chi}^\dagger_*~g^{ij}\gamma_i\left(\frac\partial{\partial
x^j}+iA_j\right)\,\chi+\chi_*^\dagger~
g_{ij}\gamma_*^iA_*^j~\widehat{\chi}\right.
\\~~~~~~~~~~~~~~~~~~~~~~~~~~~~~~~~~~~~~~~~~~
\left.-\,i\widehat{\chi}^\dagger~g_{ij}\gamma_*^i\left(\frac\partial{\partial
x_j^*}+iA_*^j\right)\,\chi_*+\chi^\dagger~
g^{ij}\gamma_iA_j~\widehat{\chi}_*\right]\end{array}}
\label{fermactionexpl}\eeq
The (left) action of gauge potentials on fermion fields in \eqn{fermactionexpl}
is given by the action of the tachyon generators
\beq
\left(V_{q^\pm}f\right)_{r^\pm}=\e^{2\pi iq_i^\pm
g^{ij}r_j^\pm}\,f_{r^\pm+q^\pm}
\label{gaugeL2action}\eeq
on functions $f\in{\cal S}(\Lambda)$. Note that \eqn{fermactionexpl} naturally
includes the (left) action of gauge potentials on anti-fermion fields.

The duality transformation is defined by interchanging starred quantities with
un-starred ones. In terms of the fields of the gauge theory this is the mapping
\beq
A_i\leftrightarrow
g_{ij}A_*^j~~~~~~,~~~~~~\chi\leftrightarrow\chi_*~~~~~~,~~~~~~\widehat{\chi}
\leftrightarrow\widehat{\chi}_*
\label{fielddualtransf}\eeq
while in terms of the geometry of the space $T_d\times T_d^*$ we have
\beq
x^i\leftrightarrow g^{ij}x^*_j~~~~~~,~~~~~~\gamma_i\leftrightarrow
g_{ij}\gamma_*^j
\label{geomdualtransf}\eeq
for all $i=1,\dots,d$. It is easily seen that both the bosonic and fermionic
actions above are invariant under this transformation, so that
\beq
I[\chi,\chi_*,\widehat{\chi},\widehat{\chi}_*;\nabla,\nabla_{\!*}]=
I[\chi_*,\chi,\widehat{\chi}_*,\widehat{\chi};\nabla_{\!*},\nabla]
\label{dualinvaction}\eeq
More general duality transformations can also be defined by
\eqn{fielddualtransf} and \eqn{geomdualtransf} (as well as the spinor
conditions \eqn{chidefs}) taken over only a subset of all the coordinate
directions $i = 1,\dots, d$. In all cases we find a manifestly duality
invariant gauge theory. In a four-dimensional spacetime, the left-right
symmetric combination $F+JFJ^{-1}$ of a field strength is relevant to the
proper addition of a topological term for the gauge field to the Yang-Mills
action yielding a gauge theory that has an explicit (anti-)self-dual form
\cite{gracia}. As we will describe in the next section, the extra terms in
\eqn{bosactionexpl} incorporate, in a certain sense to be explained, the
``dual" $J\widetilde{F}_{ij}^{\nabla,\nabla_{\!*}}J^{-1}$ to the field
strength $F_{ij}^{\nabla,\nabla_{\!*}}$. The action \eqn{bosactionexpl} is
therefore also naturally invariant under the symmetry
$F_{ij}^{\nabla,\nabla_{\!*}}\leftrightarrow
J\widetilde{F}_{ij}^{\nabla,\nabla_{\!*}}J^{-1}$ which can be thought of as a
particle-antiparticle duality on $T_d\times T_d^*$ with respect to the chiral
Lorentzian metric \eqn{chiralmetric}.

More generally, the action \eqn{totaction} is invariant under the automorphism
group given in Proposition 3. The gauge group is the affine Lie group
${\cal
P}_0\,\widehat{\rm Inn}{ }^{(0)}(\alg){\cal P}_0$ which contains the duality
symmetries and also the diffeomorphisms of $T_d\times T_d^*$ generated by the
Heisenberg fields. Thus the gauge invariance of \eqn{totaction} also naturally
incorporates the gravitational interactions in the target space. The
``diffeomorphism" invariance of the action under the group ${\cal P}_0\,{\rm
Out}(\alg){\cal P}_0$ naturally incorporates the $O(d,d;\zed)$ Morita
equivalences between classically distinct theories. The discrete group
$O(d,d;\zed)$ acts on the gauge potentials as
\beq
\pmatrix{A_i\cr A_*^i}~\mapsto~\pmatrix{(A^\top)_i^{~j}&(C^\top)_{ij}\cr
(B^\top)_i^{~j}&(D^\top)_{ij}\cr}\pmatrix{A_j\cr A_*^j}
\label{Oddaction}\eeq
where the $d\times d$ matrices $A,B,C,D$ are defined as in Proposition 2.
Again,
this symmetry is the statement that compactifications on Morita equivalent tori
are physically equivalent. The $O(2,\real)$ part of this group yields the
discrete duality symmetries described above and in general it rotates the two
gauge potentials among each other as
\beq
\pmatrix{A_i+g_{ij}A_*^j\cr
A_i-g_{ij}A_*^j}~\mapsto~\pmatrix{\cos\theta&\sin\theta\cr
-\sin\theta&\cos\theta\cr}\pmatrix{A_i+g_{ij}A_*^j\cr
A_i-g_{ij}A_*^j}~~~~~~,~~~~~~\theta\in[0,2\pi)
\label{O2action}\eeq
for each $i=1,\dots,d$. The action is in this sense a complete isomorphism
invariant of the twisted module $\widetilde{\cal T}_\omega^d$. The key feature
leading to this property is that \eqn{totaction} is a spectral invariant of the
Dirac operators \eqn{Diractachyon}.

\newsection{Physical Characteristics of $\widetilde{\cal T}_\omega^d$}

Having obtained a precise duality-symmetric characterization of the twisted
module $\widetilde{\cal T}_\omega^d$ over the noncommutative torus, we now
discuss some heuristic aspects of it using the gauge theory developed in the
previous section. We remark first of all that the operators \eqn{YMcurv} and
\eqn{YMcurvdual}, which can be interpreted as Yang-Mills curvatures, change
sign under the above duality transformation. This signals a change of
orientation of the vector bundle (represented by the finitely-generated
projective module) over the tachyon algebra. Such changes of orientation of
vector bundles under duality are a common feature of explicitly
duality-symmetric quantum field theories \cite{cheung}.

It is interesting to note that in this duality symmetric framework there are
analogs of the usual Yang-Mills instantons in {\it any} dimension. They are
defined by the curvature condition
\beq
F_{ij}^{\nabla,\nabla_{\!*}}=-J\,\widetilde{F}_{ij}^{\nabla,\nabla_{\!*}}
\,J^{-1}
\label{insteqs}\eeq
Since the bosonic action functional \eqn{bosactionexpl} is the ``square" of the
operator $F^{\nabla,\nabla_{\!*}}+J\widetilde{F}^{\nabla,\nabla_{\!*}}J^{-1}$,
the equations \eqn{insteqs} determine those gauge field configurations at which
the bosonic action functional attains its global minimum of 0. They therefore
define instanton-like solutions of the duality-symmetric gauge theory. It is in
this sense that the operator $J$ acts to map the field strength
$\widetilde{F}^{\nabla,\nabla_{\!*}}$ into the dual of
$F^{\nabla,\nabla_{\!*}}$. That the instanton charge (or Chern number) here is
0 follows from the fact that the gauge theory we constructed in section 6 was
built on a trivial vector bundle where the module of sections is the algebra
itself. To obtain instanton field configurations with non-trivial topological
charges one needs to use non-trivial bundles which are constructed using
non-trivial projectors. It would be very interesting to generalize the gauge
theory of this paper to twisted and also non-abelian modules.

Although the general solutions of the equations \eqn{insteqs} appear difficult
to deduce, there is one simple class that can be immediately identified. For
this, we consider the diagonal subgroup $\Lambda_{\rm
diag}$ of the Narain lattice \eqn{Lambdadef}, which we can decompose into two
subgroups $\Lambda_{\rm diag}^\pm$ that are generated by the bases
$\{e^i\oplus(\pm g^{ij}e_j)\}_{i=1}^d$, respectively. These rank $d$ lattices
define, respectively, self-dual and anti-self-dual $d$-dimensional tori
$T_d^\pm\equiv\real^d/2\pi\Lambda_{\rm diag}^\pm\subset T_d\times T_d^*$. Then,
on these tori, the gauge potentials obey the self-duality and anti-self-duality
conditions
\beq
x^i=\pm\,g^{ij}x_j^*~~~~~~,~~~~~~A_i=\pm\,g_{ij}A_*^j
\label{sdasdconds}\eeq
On these gauge field configurations the curvatures \eqn{YMcurv} and
\eqn{YMcurvdual} vanish identically,
\beq
F_{ij}^{\nabla,\pm\nabla}=\widetilde{F}_{ij}^{\nabla,\pm\nabla}=0
\label{curv0}\eeq
Thus the self-dual and anti-self-dual gauge field configurations
provide the analog of the instanton solutions which minimize the usual
Euclidean Yang-Mills action functional. The conditions \eqn{sdasdconds} can be
thought of as projections onto the classical sector of the theory in which
there is only a single, physical gauge potential in a $d$-dimensional
spacetime. In the classical theory duality symmetries are absent and so the
gauge field action vanishes identically. The Yang-Mills type action functional
\eqn{bosactionexpl} can therefore be thought of as measuring the amount of
asymmetry between a given connection and its dual on $\widetilde{\cal
T}_\omega^d$. As such, it measures how much duality symmetry is present in the
target space and hence how far away the stringy perturbation is from ordinary
classical spacetime. The action \eqn{bosactionexpl} can thus be regarded as an
effective measure of distance scales in spacetime.

There are other interesting physical projections of the theory that one can
make which, unlike the relations \eqn{sdasdconds}, break the duality symmetries
explicitly. For instance, consider the projection $T_d\times T_d^*\to T_d^+$
along with the freezing out of the dual gauge degrees of freedom. This
means that the fields now depend only on the local coordinates
$x^i=g^{ij}x^*_j$, and the dual gauge potential $A_*^i$ is frozen at some
constant value,
\beq
(x,x^*)~\to~(x,x)~~~~~~,~~~~~~A_*^i~\to~{\rm const.}
\label{gaugeprojn}\eeq
Since each $A_*^i$ is constant, it follows that $[A_*^i, A_*^j] = 0$. The
field strength \eqn{YMcurvdual} is then identical to \eqn{YMcurv} which
becomes the usual Yang-Mills curvature of the $d$-dimensional gauge field $A_i$
over $T_d^+$,
\beq
F_{ij}[A]~\equiv~
F_{ij}^{\nabla,\nabla_{\!*}}\Bigm|_{T_d^+}~=~-\widetilde{F}_{ij}^
{\nabla,\nabla_{\!*}}\Bigm|_{T_d^+}~=~\partial_iA_j-\partial_jA_i+i[A_i,A_j]
\label{curvprojn}\eeq
The bosonic action functional can be easily read off from \eqn{Diracdiff},
\beq
\Pi\left({\partial\slash}_\nabla^{\,2}-\overline{\partial\slash}_\nabla^{\,2}
\right)\Bigm|_{T_d^+}=-\mbox{$\frac
i4$}\left[\gamma_+^i,\gamma_-^j\right]\otimes\left(F_{ij}[A]-JF_{ij}[A]J^{-1}
\right)
\label{bosactionprojn}\eeq
When the operator \eqn{bosactionprojn} is squared, the resulting bosonic action
has the form of a symmetrized Yang-Mills functional for the gauge field $A_i$
on
$T_d^+$. It is remarkable that the projection \eqn{gaugeprojn} reduces
the bosonic action functional \eqn{bosactiondef} to the standard Yang-Mills
action used in noncommutative geometry \cite{gracia}.

In the infrared limit $g\to\infty$ ($\omega\to0$), the Moyal bracket
\eqn{moyalbracket} vanishes and the gauge theory generated by
\eqn{bosactionprojn} becomes the usual electrodynamics on the commutative
manifold $T_d^+$. Thus at large-distance scales, we recover the usual
commutative classical limit with the canonical abelian gauge theory defined on
it \cite{Book,giannibook}. The continuous ``internal" space $T_d^*$ of the
string spacetime acts to produce a sort of Kaluza-Klein mechanism by inducing
nonabelian degrees of freedom when the radii of compactification are made very
small. This nonabelian generating mechanism
is rather different in spirit than the usual ones of noncommutative geometry
which extend classical spacetime, represented by the commutative algebra
$C^\infty(T_d)$, by a discrete internal space, represented typically by a
noncommutative finite-dimensional matrix algebra. In the present case the
``internal" space comes from the natural embedding of the classical spacetime
into the noncommutative string spacetime represented by the tachyon sector of
the vertex operator algebra. In this context we find that the role of the
noncommutativity of spacetime coordinates at very short distance scales is to
induce internal (nonabelian) degrees of freedom.

As for the fermionic sector of the field theory, the projection above applied
to the spinor fields is defined as
\beq
\chi\Bigm|_{T_d^+}=\chi_*\Bigm|_{T_d^+}~~~~~~,~~~~~~\widehat{\chi}
\Bigm|_{T_d^+}=\widehat{\chi}_*\Bigm|_{T_d^+}
\label{fermprojn}\eeq
with $\gamma_i=g_{ij}\,\gamma_*^j$. Then, denoting the constant value of
$A_*^i$ by $M^i$, it follows that the fermionic action \eqn{fermactionexpl}
becomes the usual gauged Dirac action for the fermion fields
$(\widehat{\chi},\chi)$ minimally coupled to the nonabelian Yang-Mills gauge
field $A_i$ and with mass parameters $M^i$,
\beq
I_F[\chi,\widehat{\chi};\nabla,\nabla_{\!*}]\Bigm|_{T_d^+}=I_{\rm
Dirac}[\chi,\widehat{\chi},M;A]
\label{fermactionprojn}\eeq
Thus the internal symmetries of the string geometry also induce fermion masses,
and so the explicit breaking of the duality symmetries, required to project
onto physical spacetime, of the twisted module acts as a sort of geometrical
mass generating mechanism. Again in the classical limit $g\to\infty$ the left
action \eqn{gaugeL2action} becomes ordinary multiplication and
\eqn{fermactionprojn} becomes the Dirac action for $U(1)$ fermions coupled to
electrodynamics. The fact that the Kaluza-Klein modes coming from $T_d^-$
induce nonabelian degrees of freedom and fermion masses could have important
ramifications for string phenomenology. In particular, when $d=4$, the above
projections suggest a stringy origin for the canonical action of the standard
model. We remark again that the nonabelian
gauge group thus induced is the natural enhancement of the generic abelian
$U(1)^d$ gauge symmetry within the vertex operator algebra \cite{lscastro}. It
is intriguing that both nonabelian gauge degrees of freedom and masses are
induced so naturally by string geometry, and it would be interesting to study
the physical consequences of this feature in more depth.

It would also be very interesting to give a physical origin for the
duality-symmetric noncommutative gauge theory developed here using either
standard model or $M$-Theory physics. For instance, in \cite{dh} it is argued
that the characteristic interaction term for gauge theory on the noncommutative
torus, within the framework of compactified Matrix Theory \cite{cds}, appears
naturally as the worldline field theories of $N$ D-particles. This observation
follows from careful consideration of the action of $T$-duality on superstring
data in the presence of background fields. What we have shown here is that the
natural algebraic framework for the noncommutative geometry of string theory
(i.e. vertex operator algebras) has embedded within it a very
special representation of the noncommutative torus, and that the particular
module $\widetilde{\cal T}_\omega^d$ determines a gauge theory which is
manifestly duality-invariant. Thus we can derive an explicitly
duality-symmetric field theory based on very basic principles of string
geometry. This field theory is manifestly covariant, at the price of involving
highly non-local interactions. The non-locality arises from the algebraic
relations of the vertex operator algebra and as such it reflects the nature of
the string interactions. The fact that string dynamics control the very
structure of the field theory should mean that it has a more direct
relationship to $M$-Theory dynamics. One possible scenario would be to
interpret the dimension $N$ of the matrices which form the dynamical variables
of Matrix Theory as the winding numbers of strings wrapping around the
$d$-torus $T_d$. In the 11-dimensional light-cone frame, $N$ is related to the
longitudinal momentum as $p_+\propto N$, so that the vertex operator algebra is
a dual model for the $M$-Theory dynamics in which the light-cone momenta are
represented by winding numbers. The limit $N\to\infty$ of infinite winding
number corresponds to the usual Matrix Theory description of $M$-Theory
dynamics in the infinite momentum frame. The origin of the noncommutative torus
as the infinite winding modes of strings wrapping around $T_d$ is naturally
contained within the tachyon sector of the vertex operator algebra and yields a
dual picture of infinite momentum frame dynamics.

We have also seen that the relationship between lattice vertex operator
algebras and the noncommutative torus implies a new physical interpretation of
Morita equivalence in terms of target space duality transformations. It would
be interesting to investigate the relationship between this duality and the
non-classical Nahm duality which maps instantons on one noncommutative torus to
instantons on another (dual) one \cite{cds,schwarz}. It appears that the
duality interpretations in the case of the twisted module $\widetilde{\cal
T}_\omega^d$ are somewhat simpler because of the special relationship between
the deformation parameters $\omega^{ij}$ and the metric of the compactification
lattice (see Proposition 1). This relation in essence breaks some of the
symmetries
of the space. In any case, it remains to study more the gauge group
of the present module given the results of \cite{lscastro} and hence probe more
in depth the structure of the gauge theory developed here.

\bigskip

\noindent
{\bf Acknowledgements:} We thank A. Connes, S. Majid and L. Pilo for helpful
discussions. {\sc g.l.} thanks all members of DAMTP and in particular
Prof. M. Green for the kind hospitality in Cambridge.
{\sc g.l.} is a fellow of the Italian National Council of Research
(CNR) under grant CNR-NATO 215.29/01. The work of {\sc r.j.s.} was supported in
part by the Particle Physics and Astronomy Research Council (U.K.).

\newpage
\setcounter{section}{0}
\setcounter{subsection}{0}
\setcounter{equation}{0}
\renewcommand{\thesection}{Appendix \Alph{section}}
\renewcommand{\theequation}{\Alph{section}.\arabic{equation}}

\def\ce{{\cal E}}
\def\ota{\otimes_{\cal A}}
\def\bra#1{\left\langle #1\right|}
\def\ket#1{\left| #1\right\rangle}
\def\hs#1#2{\left\langle #1,#2\right\rangle}
\def\norm#1{{\Vert#1\Vert}}

\newsection{Morita Equivalence of $C^*$-algebras}

In this appendix we shall briefly describe the notion of (strong) Morita
equivalence for $C^*$-algebras \cite{Rimor}. Additional details can be found in
\cite{giannibook}, for example. Throughout this appendix $\alg$ is an
arbitrary unital $C^*$-algebra whose norm we denote by $\norm{\cdot}$.

A right Hilbert module over $\alg$ is a right $\alg$-module $\ce$ \footnote{In
more simplistic terms this means that $\ce$ carries a right action of $\alg$.}
endowed with an $\alg$-valued Hermitian structure, i.e. a sesquilinear form
$\hs{\cdot~}{~\cdot}_\alg : \ce \times \ce \rightarrow \alg$
which is conjugate linear in its first argument and satisfies
\bea
\hs{\eta_1}{\eta_2 a}_\alg& = &\hs{\eta_1}{\eta_2}_\alg a\label{hsp1}\\
\hs{\eta_1}{\eta_2}_\alg^*& = &\hs{\eta_2}{\eta_1}_\alg\label{hsp2}\\
\hs{\eta}{\eta}_\alg \geq 0~~&,& ~~\hs{\eta}{\eta}_\alg = 0~
\Leftrightarrow~\eta = 0
\label{hsp3}\eea
for all $\eta_1, \eta_2, \eta \in \ce$, $a \in \alg$. We define a norm on $\ce$
by $\norm{\eta}_\alg = \sqrt{\norm{\hs{\eta}{\eta}_\alg}}$ for any $\eta
\in\ce$ and require that $\ce$ be complete with respect to this norm. We also
demand that the module be full, i.e. that the ideal ${\rm
span}_\complexs\{\hs{\eta_1}{\eta_2}_\alg~|~\eta_1, \eta_2 \in \ce\}$ is dense
in $\alg$ with respect to the norm closure. A left Hilbert module structure on
a left $\alg$-module $\ce$ is provided by an $\alg$-valued Hermitian structure
$\hs{\cdot~}{~\cdot}_\alg$ on $\ce$ which is conjugate linear in its second
argument and with the condition \eqn{hsp1} replaced by
\beq
\hs{a\eta_1}{\eta_2}_\alg = a\hs{\eta_1}{\eta_2}_\alg~~~~~~\forall
\eta_1,\eta_2\in \ce, ~a \in \alg
\label{hspl}\eeq

Given a Hilbert module $\ce$, its compact endomorphisms are obtained as usual
from the `endomorphisms of finite rank'. For any $\eta_1, \eta_2 \in \ce$ an
endomorphism
$\ket{\eta_1}\bra{\eta_2}$ of $\ce$ is defined by
\beq
\label{bk}
\ket{\eta_1}\bra{\eta_2} (\xi) = \eta_1 \hs{\eta_2}{\xi}_\alg~~~~,
{}~~~~\forall\xi \in\ce~
\eeq
which is right $\alg$-linear,
\beq
\ket{\eta_1}\bra{\eta_2} (\xi a) = \Bigl(\ket{\eta_1}\bra{\eta_2}
(\xi)\Bigr) a~~~~,
{}~~~~\forall\xi \in \ce,~ a \in \alg
\eeq
Its adjoint endomorphism is given by
\beq
\Bigl(\ket{\eta_1}\bra{\eta_2}\Bigr)^* =
\ket{\eta_2}\bra{\eta_1}~~~~,~~~~\forall\eta_1, \eta_2 \in \ce
\eeq
It can be shown that for $\eta_1, \eta_2, \xi_1, \xi_2 \in \ce$ one has the
expected composition rule
\beq
\ket{\eta_1}\bra{\eta_2} \circ \ket{\xi_1}\bra{\xi_2} =
\ket{\eta_1 \hs{\eta_2}{\xi_1}_\alg} \bra{\xi_2}
= \ket{\eta_1} \bra{ \hs{\eta_2}{\xi_1}_\alg \xi_2}
\eeq

It turns out that if $\ce$ is a finitely-generated projective module then the
norm closure ${\rm End}^0_\alg(\ce)$ (with respect to the natural operator norm
which yields a $C^*$-algebra) of the $\complex$-linear span of the
endomorphisms of the form \eqn{bk} coincides with the endomorphism algebra
${\rm End}_\alg(\ce)$ of $\ce$. In fact, this property completely characterizes
finitely-generated projective
modules. For then, there are two finite sequences $\{\xi_k\}$ and $\{\zeta_k\}$
of elements of $\ce$ such that the identity endomorphism $\id_\ce$ can be
written as $\id_\ce = \sum_k \ket{\xi_k} \bra{\zeta_k}$. For any $\eta \in
\ce$, we then have
\beq
\eta = \id_\ce\,\eta = \sum_k \ket{\xi_k} \bra{\zeta_k}\eta = \sum_k \xi_k
\hs{\zeta_k}{\eta}_\alg
\eeq
and hence $\ce$ is finitely-generated by the sequence $\{\xi_k\}$. If $N$ is
the
length of the sequences $\{\xi_k\}$ and $\{\zeta_k\}$,
we can embed $\ce$ as a direct summand of $\alg^N\equiv\bigoplus_{n=1}^N\alg$,
proving that it is projective. The embedding and surjection maps are
defined, respectively, by
\bea
&& \lambda : \ce \rightarrow \alg^N~~~~, ~~~~ \lambda(\eta)
= \Bigl(\hs{\zeta_1}{\eta}_\alg, \dots, \hs{\zeta_N}{\eta}_\alg\Bigr)
\nonumber \\&& \rho : \alg^N \rightarrow \ce~~~~, ~~~~ \rho\Bigl((a_1, \dots,
a_N)\Bigr) = \sum_k \xi_k a_k
\eea
Then, given any $\eta \in \ce$, we have
\beq
\rho \circ \lambda (\eta) = \rho\Bigl(\left(\hs{\zeta_1}{\eta}_\alg, \dots,
\hs{\zeta_N}{\eta}_\alg\right)\Bigr) = \sum_k \xi_k \hs{\zeta_k}{\eta}_\alg
= \sum_k \ket{\xi_k} \bra{\zeta_k}(\eta)=\id_\ce(\eta)
\eeq
so that $\rho \circ \lambda = \id_\ce$, as required. The projector $p = \lambda
\circ \rho$ identifies $\ce$ as $p\alg^N$.

For any full Hilbert module $\ce$ over a $C^*$-algebra $\alg$, the
latter is {\it (strongly) Morita equivalent} to the $C^*$-algebra ${\rm
End}^0_\alg(\ce)$ of compact endomorphisms of $\ce$. If $\ce$ is
finitely-generated and projective, so that ${\rm End}^0_\alg(\ce) ={\rm
End}_\alg(\ce)$, then the algebra $\alg$ is strongly Morita equivalent to the
whole of ${\rm End}_\alg(\ce)$. The equivalence is expressed as follows. The
idea is to construct an ${\rm End}^0_\alg(\ce)$-valued Hermitian structure on
$\ce$ which is compatible with the Hermitian structure
$\hs{\cdot~}{~\cdot}_\alg$. Consider then a full {\it right} Hilbert module
$\ce$ over the algebra $\alg$ with $\alg$-valued Hermitian structure
$\hs{\cdot~}{~\cdot}_\alg$. It follows that $\ce$ is a {\it left} module
over the $C^*$-algebra ${\rm End}^0_\alg(\ce)$. A left Hilbert module structure
is constructed by inverting the definition \eqn{bk} so as to produce an ${\rm
End}^0_\alg(\ce)$-valued Hermitian structure on $\ce$,
\beq
\hs{\eta_1}{\eta_2}_{{\rm End}^0_\alg(\ce)} = \ket{\eta_1}\bra{\eta_2}~~~~,
{}~~~~\forall\eta_1, \eta_2 \in \ce
\label{hscom}\eeq
It is straightforward to check that \eqn{hscom} satisfies all the properties of
a left Hermitian structure including conjugate linearity in its second
argument. It follows from the definition of a compact endomorphism that the
module $\ce$ is also full as a module over ${\rm End}^0_\alg(\ce)$.
Furthermore, from the definition \eqn{bk} we have a compatibility condition
between the two Hermitian structures on $\ce$,
\beq
\hs{\eta_1}{\eta_2}_{{\rm End}^0_\alg(\ce)} \xi = \ket{\eta_1}\bra{\eta_2}(\xi)
=\eta_1 \hs{\eta_2}{\xi}_\alg~~~~, ~~~~\forall\eta_1, \eta_2, \xi \in \ce
\eeq
The Morita equivalence is also expressed by saying that the module $\ce$
is an ${\rm End}^0_\alg(\ce)$-$\alg$ equivalence Hilbert bimodule.

A $C^*$-algebra ${\cal B}$ is said to be Morita equivalent to the $C^*$-algebra
$\alg$ if ${\cal B} \cong {\rm End}^0_\alg(\ce)$ for some
$\alg$-module $\ce$. Morita equivalent $C^*$-algebras have equivalent
representation theories. If the $C^*$-algebras $\alg$ and ${\cal B}$ are Morita
equivalent with ${\cal B}$-$\alg$ equivalence bimodule $\ce$, then given a
representation of $\alg$, using $\ce$ we can construct a unitary equivalent
representation of ${\cal B}$. For this, let $({\cal H}, \pi_\alg)$ be a
representation of $\alg$ on a Hilbert space ${\cal H}$. The algebra
$\alg$ acts by bounded operators on the left on ${\cal H}$ via $\pi_\alg$.
This action can be used to construct another Hilbert space
\beq
{\cal H}' = \ce \ota {\cal H}~~~~,~ ~~~\eta a \ota \psi - \eta
\ota \pi_\alg(a) \psi = 0~~,~~\forall a\in\alg, ~\eta\in\ce,~\psi\in{\cal H}
\eeq
with scalar product
\beq
(\eta_1 \ota \psi_1, \eta_2 \ota \psi_2)_{{\cal H}'} = (\psi_1,
\hs{\eta_1}{\eta_2}_\alg \psi_2)_{\cal H}~~~~,
{}~~~~\forall\eta_1, \eta_2 \in\ce, ~\psi_1, \psi_2 \in {\cal H}
\eeq
A representation $({\cal H}', \pi_{\cal B})$ of the algebra ${\cal B}$ is
then defined by
\beq
\pi_{\cal B}(b)(\eta \ota \psi) = (b \eta) \ota \psi~~~~,~~~~\forall
b\in{\cal B},~\eta \ota \psi\in{\cal H}'
\eeq
This representation is unitary equivalent to the representation
$({\cal H}, \pi_\alg)$. Conversely, starting with a representation of
${\cal B}$, we can use a conjugate $\alg$-${\cal B}$ equivalence bimodule
$\overline{\ce}$ to construct an equivalent representation of $\alg$. Morita
equivalence also yields isomorphic $K$-groups and cyclic homology, so that
Morita equivalent algebras determine the same noncommutative geometry. However,
the physical characteristics can be drastically different. For example, the
algebras can have different (unitary) gauge groups and hence determine
physically inequivalent gauge theories.


\newpage

\begin{thebibliography}{99}

\baselineskip=12pt

\bibitem{Rieffel} M.A. Rieffel, {\it $C^*$-algebras associated with Irrational
Rotations}, Pac. J. Math. {\bf 93} (1981) 415.

\bibitem{ncmodules} A. Connes, {\it $C^*$-alg\`{e}bres et G\'{e}om\'{e}tries
Diff\'{e}rentielle}, C. R. Acad. Sci. Paris {\bf A290} (1980) 599;\\ A. Connes
and M.A. Rieffel, {\it Yang-Mills for Noncommutative Two-tori}, Contemp. Math.
{\bf 62} (1987) 237;\\ M.A. Rieffel, {\it Projective Modules over
Higher-dimensional Noncommutative Tori}, Can. J. Math. {\bf 40} (1988) 257.

\bibitem{Rieffel1} M.A. Rieffel, {\it The Cancellation Theorem for Projective
Modules over Irrational Rotation $C^*$-algebras}, Proc. London Math. Soc. {\bf
47} (1983) 285.

\bibitem{Book} A.~Connes, {\it Noncommutative Geometry} (Academic Press,
1994).

\bibitem{giannibook} G. Landi, {\it An Introduction to Noncommutative Spaces
and their Geometries} (Springer, 1997).

\bibitem{cds} A. Connes, M.R. Douglas and A. Schwarz, {\it Noncommutative
Geometry and Matrix Theory: Compactification on Tori}, J. High Energy Phys.
{\bf 9802} (1998) 003.

\bibitem{dh} M.R. Douglas and C.M. Hull, {\it $D$-branes and the Noncommutative
Torus}, J. High Energy Phys. {\bf 9802} (1998) 008;\\ M. Li, {\it Comments on
Supersymmetric
Yang-Mills Theory on a Noncommutative Torus}, hep-th/9802052;\\ M. Berkooz,
{\it Nonlocal Field Field Theories and the Noncommutative Torus}, Phys.\
Lett.\ {\bf B430} (1998) 237;\\ Y.-K.E. Cheung and M. Krogh, {\it
Noncommutative Geometry from 0-branes in a Background $B$-field}, Nucl.\ Phys.
{\bf B528} (1998) 185;\\ D. Bigatti, {\it
Noncommutative Geometry and Super Yang-Mills Theory}, hep-th/9804120.

\bibitem{nctmatrix} P.-M. Ho, Y.-Y. Wu and Y.-S. Wu, {\it Towards a
Noncommutative Geometric Approach to Matrix Compactification},
Phys.\ Rev.\ {\bf D58} (1998) 026006;\\ T. Kawano and K. Okuyama, {\it Matrix
Theory on Noncommutative Torus}, Phys. Lett. {\bf B433} (1998) 29--34;\\ F.
Ardalan, H. Arfaei and M.M.
Sheikh-Jabbari, {\it Mixed Branes and Matrix Theory on Noncommutative Torus},
hep-th/9803067;\\ P.-M. Ho, {\it Twisted Bundle on Quantum Torus and BPS States
in Matrix Theory}, Phys. Lett. {\bf B434} (1998) 41--47.

\bibitem{Matrixmodule} P.-M. Ho and Y.-S. Wu, {\it Noncommutative Gauge
Theories in Matrix Theory}, Phys.\ Rev.\ {\bf D58} (1998) 066003;\\ R.
Casalbuoni, {\it Algebraic Treatment of Compactification on Noncommutative
Tori}, Phys. Lett. {\bf B431} (1998) 69.

\bibitem{Dfield} E. Witten, {\it Bound States of Strings and $p$-branes}, Nucl.
Phys. {\bf B460} (1996) 335;\\ M.R. Douglas, D. Kabat, P. Pouliot and S.H.
Shenker, {\it $D$-branes and Short Distances in String Theory}, Nucl. Phys.
{\bf B485} (1997) 85--127.

\bibitem{howu} P.-M. Ho and Y.-S. Wu, {\it Noncommutative Geometry and
$D$-branes}, Phys. Lett. {\bf B398} (1997) 52--60.

\bibitem{lms} F. Lizzi, N.E. Mavromatos and R.J. Szabo, {\it Matrix
$\sigma$-models for Multi $D$-brane Dynamics}, Mod. Phys. Lett. {\bf A13}
(1998) 829--842.

\bibitem{Matrix} T. Banks, W. Fischler, S.H. Shenker and L. Susskind, {\it M
Theory as a Matrix Model: A Conjecture}, Phys. Rev. {\bf D55} (1997) 5112.

\bibitem{fg} J. Fr\"{o}hlich and K. Gaw\c edzki, {\it Conformal Field Theory
and Geometry of Strings}, CRM Proc. Lecture Notes {\bf 7} (1994) 57.

\bibitem{Cham} A.H. Chamseddine, {\it The Spectral Action Principle in
Noncommutative Geometry and the Superstring}, Phys. Lett. {\bf B400} (1997) 87;
{\it An Effective Superstring Spectral Action}, Phys. Rev. {\bf D56} (1997)
3555.

\bibitem{lslett} F. Lizzi and R.J. Szabo, {\it Target Space Duality in
Noncommutative Geometry}, Phys. Rev. Lett. {\bf 79} (1997) 3581.

\bibitem{lslong} F. Lizzi and R.J. Szabo, {\it Duality Symmetries and
Noncommutative Geometry of String Spacetimes}, Commun. Math. Phys. {\bf 197}
(1998) 667.

\bibitem{Morita} M.A. Rieffel and A. Schwarz, {\it Morita Equivalence of
Multidimensional Noncommutative Tori}, math.QA/9803057, to appear in Int.
J. of Math.

\bibitem{schwarz} A. Schwarz, {\it Morita Equivalence and Duality}, Nucl. Phys.
{\bf B534} (1998) 720--738.

\bibitem{lsem} F. Lizzi and R.J. Szabo, {\it Electric-magnetic Duality in
Noncommutative Geometry}, Phys. Lett. {\bf B417} (1998) 303.

\bibitem{lscastro} F. Lizzi and R.J. Szabo, {\it Noncommutative Geometry and
Spacetime Gauge Symmetries of String Theory}, Chaos, Solitons and Fractals {\bf
10} (1999) 445.

\bibitem{faddeev} L.D. Faddeev, {\it Discrete Heisenberg-Weyl Group and
Modular Group}, Lett. Math. Phys. {\bf 34} (1995) 249.

\bibitem{sm} A. Connes and J. Lott, {\it Particle Models and Noncommutative
Geometry}, Nucl. Phys. B (Proc. Suppl.) {\bf 18B} (1990) 29;\\ C.P.
Mart\'{\i}n, J.M. Gracia-Bond\'{\i}a and J.C. V\'arilly, {\it The Standard
Model as a Noncommutative Geometry: The Low Energy Regime}, Phys. Rep. {\bf
294} (1998) 363.

\bibitem{connesreal} A. Connes, {\it Noncommutative Geometry and Reality}, J.
Math. Phys. {\bf 36} (1995) 619.

\bibitem{connesauto} A. Connes, {\it Gravity Coupled with Matter and the
Foundation of Noncommutative Geometry}, Commun. Math. Phys. {\bf 182} (1996)
155.

\bibitem{specaction} A.H. Chamseddine and A. Connes, {\it Universal Formula for
Noncommutative Geometry Actions: Unification of Gravity and the Standard
Model}, Phys. Rev. Lett. {\bf 77} (1996) 4868; {\it The Spectral Action
Principle}, Commun. Math. Phys. {\bf 186} (1997) 731.

\bibitem{zwan} D. Zwanziger, Phys. Rev. {\bf D3} (1971) 880;\\ S. Deser and C.
Teitelboim, {\it Duality Transformations of Abelian and Nonabelian Gauge
Fields}, Phys. Rev. {\bf D13} (1976) 1592--1597;\\ J.H. Schwarz and A. Sen,
{\it Duality Symmetric Actions}, Nucl. Phys. {\bf B411} (1994) 35;\\ S. Deser,
A. Gomberoff, M. Henneaux and C. Teitelboim, {\it Duality, Selfduality, Sources
and Charge Quantization in Abelian $N$-form Theories}, Phys. Lett. {\bf B400}
(1997) 80.

\bibitem{deserteit} P. Pasti, D. Sorokin and M. Tonin, {\it Duality Symmetric
Actions with Manifest Spacetime Symmetries}, Phys. Rev. {\bf D52} (1995)
4277;\\ N. Berkovits, {\it Manifest Electromagnetic Duality in Closed
Superstring Field Theory}, Phys. Lett. {\bf B388} (1996) 743; {\it Local
Actions with Electric and Magnetic Sources}, {\bf B395} (1997) 28; {\it
Super-Maxwell Actions with Manifest Duality}, {\bf B398} (1997) 79;\\ R. Medina
and N. Berkovits, {\it Pasti-Sorokin-Tonin Actions in the Presence of Sources},
Phys. Rev. {\bf D56} (1997) 6388--6390.

\bibitem{black} S. Deser, M. Henneaux and C. Teitelboim, {\it Electric-magnetic
Black Hole Duality}, Phys. Rev. {\bf D55} (1997) 826;\\ D.D. Song and R.J.
Szabo, {\it Black String Entropy from Anomalous $D$-brane Couplings},
hep-th/9805027.

\bibitem{agan} M. Aganagic, J. Park, C. Popescu and J.H. Schwarz, {\it Dual
$D$-brane Actions}, Nucl. Phys. {\bf B496} (1997) 215--230;\\ A. Nurmagambetov,
{\it Duality-symmetric Three-brane and its Coupling to Type IIB Supergravity},
Phys. Lett. {\bf B436} (1998) 289.

\bibitem{cheung} Y.-K.E. Cheung and Z. Yin, {\it Anomalies, Branes and
Currents}, Nucl. Phys. {\bf B517} (1998) 69.

\bibitem{flm} I.B. Frenkel, J. Lepowsky and A. Meurman, {\it Vertex Operator
Algebras and the Monster}, Pure Appl. Math. {\bf 134} (Academic Press, New
York, 1988);\\ R.W. Gebert, {\it Introduction to Vertex Algebras, Borcherds
Algebras and the Monster Lie Algebra}, Intern. J. Mod. Phys. {\bf A8} (1993)
5441.

\bibitem{gsw} M.B. Green, J.H. Schwarz and E. Witten, {\it Superstring Theory},
(Cambridge University Press, 1987).

\bibitem{ConstScharf} F.\ Constantinescu and G.~Scharf, {\em Smeared and
Unsmeared Chiral Vertex Operators}, Commun. Math. Phys. {\bf 200} (1999)
275--296.

\bibitem{debievre} S. De Bi\`evre, {\it Chaos, Quantization and the Classical
Limit on the Torus}, Proc. XIV Workshop on {\sl Geometrical Methods in
Physics}, July 1995, Bialowieza, Poland, mp-arc 96--191;\\ J.C. V\'arilly, {\it
An Introduction to Noncommutative Geometry}, Lectures at the EMS Summer School
on {\sl Noncommutative Geometry and Applications}, September 1997, Portugal,
physics/9709045.

\bibitem{Fairlie} D. Fairlie, P. Fletcher, and C. Zachos,
{\em Trigonometric Structure Constants for New Infinite Dimensional Algebras,
} Phys. Lett. {\bf B218} (1989) 203--206; {\em Infinite-Dimensional Algebras
and a Trigonometric Basis for the Classical Lie Algebras},
J.\ Math.\  Phys. {\bf 31} (1990) 1088--1094;\\ D. Fairlie and C. Zachos, {\em
Infinite-Dimensional Algebras, Sine Brackets, and SU($\infty$)}, Phys. Lett.
{\bf B224} (1989) 101--107.

\bibitem{gracia} J.M. Gracia-Bond\'{\i}a, B. Iochum and T. Sch\"ucker, {\it The
Standard Model in Noncommutative Geometry and Fermion Doubling}, Phys. Lett.
{\bf B416} (1998) 123.

\bibitem{Rimor} M.A. Rieffel, {\it Induced Representations of $C^*$-algebras},
Bull. Amer. Math. Soc. {\bf 78} (1972) 606--609; Adv. Math. {\bf 13} (1974)
176--257; {\it Morita Equivalence for Operator Algebras}, in: {\sl Operator
Algebras and Applications}, Proc. Symp. Pure Math. {\bf 38}, R.V. Kadison ed.
(American Mathematical Society, 1982) 285--298.

\end{thebibliography}
\end{document}